\documentclass[12pt,preprint]{aastex}
\begin{document}

\parindent=1.0cm

\title{The Disk and Extraplanar Environment of NGC 247 \altaffilmark{1} \altaffilmark{2} \altaffilmark{3}}

\author{T. J. Davidge \altaffilmark{4}}

\affil{Herzberg Institute of Astrophysics,
\\National Research Council of Canada, 5071 West Saanich Road,
\\Victoria, B.C. Canada V9E 2E7\\ {\it email: tim.davidge@nrc.ca}}

\altaffiltext{1}{Based on observations obtained with MegaPrime/MegaCam, a joint 
project of the CFHT and CEA/DAPNIA, at the Canada-France-Hawaii
Telescope, which is operated by the National Research Council of Canada,
the Centre National de la Recherche Scientifique, and the University of
Hawaii.} 

\altaffiltext{2}{Based on observations obtained at the
Gemini Observatory, which is operated by the Association of Universities
for Research in Astronomy, Inc., under a co-operative agreement with the
NSF on behalf of the Gemini partnership: the National Science Foundation
(United States), the Particle Physics and Astronomy Research Council
(United Kingdom), the National Research Council of Canada (Canada),
CONICYT (Chile), the Australian Research Council (Australia), CNPq (Brazil),
and CONICET (Argentina).} 

\altaffiltext{3}{This publication makes use of data products 
from the Two Micron All Sky Survey, which is a joint project of the University of 
Massachusetts and the Infrared Processing and Analysis Center/California Institute of 
Technology, funded by the National Aeronautics and Space Administration and the 
National Science Foundation.}

\altaffiltext{4}{Visiting Astronomer, Canada-France-Hawaii
Telescope, which is operated by the National Research Council of Canada,
the Centre National de la Recherche Scientifique, and the University of
Hawaii.}

\begin{abstract}

	Images spanning the 0.5 to $2.3\mu$m wavelength interval are used to probe the 
stellar content in the disk and halo of the Sculptor Group galaxy NGC 247. The main 
sequence turn-off (MSTO) in the inner 12 kpc of the disk corresponds to an age of $\sim 
6$ Myr. A mean star formation rate (SFR) of 0.1 M$_{\odot}$ year$^{-1}$ during the past 
16 Myr is computed from star counts; this is qualitatively consistent with the relatively 
low SFR computed from other indicators. The color of the red supergiant plume does
not change with radius, suggesting that the mean metallicity of young stars does 
not vary by more than $\sim 0.1$ dex throughout the 
disk. However, the number of bright main sequence stars per 
local stellar mass density climbs towards larger radii out to a 
distance of 12 kpc; the scale lengths that characterize the radial distributions of 
young and old stars thus differ throughout much of the disk. 
For comparison, the density of bright main sequence stars with respect to 
projected HI mass gradually drops with increasing radius. 
At distances in excess of 12 kpc the population of very young stars 
disappears; the MSTO at galactocentric radii between 12 and 15 kpc corresponds to $\sim 
16$ Myr, while between 15 and 18 kpc the MSTO age is $\geq 40$ Myr. 
Stars evolving on the red giant branch are resolved in an extraplanar region that has
a projected galactocentric distance along the minor axis of $\sim 12$ kpc. 
The $r'-i'$ colors of the RGB population indicate 
that there is a broad spread in metallicity, with a mean [M/H] $\sim -1.2$. 
The RGB-tip occurs at $i' = 24.5 \pm 0.1$, and a distance modulus of $27.9 \pm 
0.1$ is computed. Luminous AGB stars are also seen in this field, and the brightnesses of 
these objects is consistent with an age $\sim 3$ Gyr. Possible origins of such a 
population of intermediate aged stars well off of the disk plane are discussed.

\end{abstract}


\section{INTRODUCTION}

	Disk galaxies have played an important role in the evolution of the Universe, 
including the earliest epochs of galaxy formation. The first proto-galactic systems likely 
obtained angular momentum from the torquing action of 
nearby mass concentrations (e.g. White 1984; Barnes \& Efstathiou 1987), and so may have 
been disks. These early disks were almost certainly obliterated in mergers, or 
structurally altered by tidal harrassment (Moore, Lake, \& Katz 1998) and interactions 
(e.g. Mayer et al. 2001; Pasetto, Chiosi, \& Carraro 2003). In the particular case of the 
Milky-Way, simulations predict that the Galactic bulge and halo formed during the merger 
of such proto-galactic disks (e.g. Bekki \& Chiba 2001; Governato et al. 2004). 

	It is not clear when the stable disk systems 
that are seen at the present day formed. 
It has been suggested that the long-lived disks in the local Universe may have 
formed relatively late, after the frequency of mergers and interactions subsided to a 
point where gas could cool and form stable structures (e.g. Weil, Eke, \& Efstathiou 1998, 
Bekki \& Chiba 2001). Such delayed disk formation helps to relieve the problem of 
over-cooling (e.g. Navarro \& Steinmetz 1997, but see also Governato et al. 2004). 
However, there is a potential conflict with observations, as 
large numbers of objects with disk-like light profiles 
populate the universe at high redshifts (e.g. Marleau \& Simard 1998; 
Labbe et al. 2003), suggesting that disks survive for long periods 
of time during the epoch assocated with the assembly of large galaxies. 

	Studies of nearby spiral galaxies appear to support the notion that 
stable disks formed at moderately high redshifts. Some of the globular cluster 
systems associated with nearby spiral galaxies have disk-like kinematics, and 
disk-related globular clusters in the LMC, M31, and Sculptor Group spirals appear to 
have old ages (Schommer et al. 1992; Morrison et al. 2004; Olsen et al. 
2004). There are some exceptions; for example, the M33 globular cluster system 
appears to be dominated by objects that formed during intermediate epochs 
(Sarajedini et al. 1998), and a population of blue globular clusters may also be 
present in NGC 2403 (Battistini et al. 1984). 
Moreover, some objects interpreted to be old, metal-poor disk 
globular clusters may actually be relatively young clusters, 
that have been mis-classified due to the age-metallicity degeneracy (Beasley et al. 
2004). As for the Galactic disk, Salaris, Weiss, 
\& Percival (2004) conclude from the ages of open clusters that the disk formed 
$\sim 2$ Gyr after the onset of star formation in the Galactic halo, while dynamical 
arguements give an age for the disk in the solar neighborhood that is only $\sim 1$ 
Gyr younger than that of typical globular clusters (Binney, Dehnen, \& Bertelli 2000).

	Studies of the stellar contents of nearby spiral galaxies will provide insight 
into the age and evolution of disks. As one of the nearest ensembles 
of late-type spiral galaxies external to the Local 
Group, the so-called Sculptor Group is an important laboratory for probing disks. 
The classical notion of a single Sculptor Group may 
not be correct, as the galaxies in this part of the sky appear to belong to 
three distinct concentrations, that are dominated by NGC 300, NGC 7793, and 
NGC 253. These concentrations are distributed over a linear distance 
in excess of 2 Mpc (Karachentsev et al. 2003), and appear 
to follow the free Hubble flow; consequently, they may not be gravitationally bound 
(Karachentsev et al. 2003). Nevertheless, the angular momentum vectors of 
Sculptor Group galaxies tend to be aligned, suggesting a common origin of angular 
momentum and a shared evolutionary heritage during early epochs (Whiting 1999). 

	In the present study, deep images obtained with the CFHT and Gemini South 
telescopes, spanning the $0.5 - 2.3\mu$m wavelength interval, are used to 
probe the bright stellar content of the Sculptor Group galaxy NGC 247, which 
is in the sub-group dominated by NGC 253. This sub-group has a crossing time of 6.9 Gyr 
(Karachentsev et al. 2003), and so is dynamically unevolved. The integrated 
brightness of NGC 247 is M$_B = -18$ (Carignan 1985), and the galaxy has been 
assigned morphological types Sc (Sandage \& Tammann 1987) and Sd 
(de Vaucouleurs et al. 1991). The disk of NGC 247 is 
inclined at an angle of 74 degrees (Carignan \& Puche 1990), allowing the 
stellar contents of the disk and extraplanar regions to be decoupled. 

	Carignan \& Puche (1990) find that the HI envelope of NGC 247 is relatively 
compact when compared with other systems, as it was detected only out to the limits of the 
stellar disk. Ferguson et al. (1996) used H$\alpha$ images to investigate the 
distribution of ionized gas in NGC 247. The H$\alpha$ emission is concentrated in 
two spiral arms, and the total H$\alpha$ flux is modest when 
compared with other spiral and irregular galaxies. Strassle et al. 
(1999) find that NGC 247 is embedded in an X-ray halo 
that accounts for roughly 0.5\% of the total galaxy mass. 
The X-ray halo has a core radius of 4.4 kpc and 
is more concentrated than the dark matter halo, 
leading Strassle et al. (1999) to suggest that there could 
be infall of material into the central regions of the galaxy. However, 
the nucleus of NGC 247 has a $J-K$ color that is similar to that of the nucleus of M33 
(Davidge \& Courteau 2002), which most recently formed stars $\sim 70$ Myr in the past 
(e.g. Gordon et al. 1999; Davidge 2000); therefore, if there is infall of gas then it 
evidently has not lead to the triggering of very recent nuclear star formation.

	Despite being relatively close, surprisingly little 
is known about the resolved stellar content of NGC 247. The only published 
study with modern detectors is that by Davidge \& Courteau (2002), who 
investigated the bright stellar content in the innermost region of the disk. In the 
current paper the stellar content of NGC 247 throughout the disk and in 
the envelope and halo is investigated. One of the datasets -- that obtained with 
the CFHT MegaPrime -- covers virtually the entire galaxy, and thereby provides for 
a complete census of the bright stellar content. The key goals 
of this study are (1) to probe the relative 
distribution of young stars with respect to the local integrated stellar and HI 
contents, and thereby search for age gradients and compare the locations of 
recent star formation and gas, (2) to estimate the linear extent of the 
stellar disk, using resolved stars as a tracer, and (3) to measure the 
age and metallicity of stars in the extraplanar regions of NGC 247.

	The paper is structured as follows. Details of the observations and the 
procedures used to reduce the data are described in \S 2, while the 
photometric measurements are discussed in \S 3. The stellar content in the HI disk
is investigated in \S 4, while the stellar content in the 
extraplanar regions and the portion of the 
disk that falls outside of the detected HI distribution is the subject of 
\S 5. A discussion and summary of the results follows in \S 6.

\section{OBSERVATIONS \& DATA REDUCTION}

\subsection{MegaPrime Data}

	Images of NGC 247 and its surroundings were recorded through $g'$, 
$r'$, and $i'$ filters with the CFHT MegaPrime imaging system. 
The detector in MegaPrime is MegaCam, which is a mosaic of thirty six $2048 \times 4612$ 
pixel CCDs deployed in a $4 \times 9$ pattern. The spatial sampling of 0.19 arcsec 
pixel$^{-1}$ permits a field that is roughly 1 degree$^2$ to be imaged with each exposure.
A comprehensive description of MegaCam has been given by Boulade et al. (2003). 

	The data were obtained on the night of UT December 23, 2003, as part of the 
queue-scheduled observing program for MegaPrime. 
Five 200 sec exposures were recorded through each filter with the galaxy 
centered on the detector mosaic. Detector element CCD03 was not working on this night; 
however, CCD03 is well separated from the portion of the mosaic that samples NGC 247, 
and so the scientific results of this program were not affected.

	The initial reduction of the data was performed with the Elixir pipeline at CFHT. 
This processing included the normalization of detector-to-detector gain variations, bias 
subtraction, the division by a twilight flat-field frame, and the subtraction of a 
fringe frame from the $i'$ data. Additional processing was done 
at HIA Victoria, where the Elixir-processed images were spatially aligned 
and median-combined. The combined images for each detector element were then trimmed to 
the area of common coverage for all exposures. Stars in the final images have 
FWHM $\sim 0.9$ arcsec. The $r'$ images of the six detector 
elements that cover the disk of NGC 247 and its immediate surroundings, 
which are the subject of the current investigation, are shown in Figure 1. 
 
\subsection{CFHTIR Data}

	Three fields in NGC 247 were observed with 
the CFHTIR imager, which was mounted at the Cassegrain focus of the CFHT. The detector in 
the CFHTIR is a $1024 \times 1024$ HgCdTe array. With an image scale of 0.21 arcsec 
pixel$^{-1}$, each exposure covers $3.6 \times 3.6$ arcmin$^2$.

	The CFHTIR data sample portions of the NGC 247 disk as well as the galaxy center, 
which was observed previously in the near-infrared by Davidge \& Courteau (2002).
The central co-ordinates of the fields are listed in 
Table 1, while the areas sampled with the CFHTIR are indicated in Figure 1. 
The data were recorded through $J, H,$ and $K'$ filters on the nights of 
UT November 22/23 and 23/24 2002. A $2 \times 2$ square dither pattern was employed, 
and the total exposure time was 12 minutes filter$^{-1}$ field$^{-1}$.

	The data were processed using a standard pipeline for near-infrared imaging. The 
basic steps in the reduction sequence are: (1) the subtraction of a dark frame, (2) the 
division by a dome flat, which had thermal emission signatures removed by subtracting 
images of the dome spot with flat-field and dome lights `off' from those 
taken with flat-field lights `on', (3) the subtraction of the 
DC sky level from each image, and (4) the subtraction of 
thermal signatures produced by warm objects along the light path. 
The calibration frames used in the last step were constructed from 
flat-fielded and DC sky-subtracted images of areas of the sky having low stellar density; 
these frames also monitor the incidence of interference fringes.

	The processed images were spatially aligned 
and then median combined to suppress cosmic-rays and bad pixels. 
As a final processing step, the combined images were trimmed to 
the region that is common to all four dither positions. Stars in 
the final processed images have FWHM between 0.8 and 1.0 arcsec.

\subsection{Gemini GMOS Data}

	Two overlapping fields that sample the outer regions of NGC 247 along the 
western side of the minor axis were observed with the GMOS on the Gemini South telescope. 
The detector in GMOS is a mosaic of three $2048 \times 4068$ EEV CCDs. Each pixel samples 
0.072 arcsec, and an area that is 5.5 arc minutes on a side is imaged on the sky. 
The raw pixels were binned $2 \times 2$ during read-out 
to better match the image sampling to the seeing. Crampton et al. (2000) describe the 
various GMOS components in detail.

	The central co-ordinates of the GMOS fields are listed in Table 1, while their 
locations are marked in Figure 1. A control field, 
located midway between NGC 55 and NGC 247 and 
discussed previously by Davidge (2005) in his investigation of NGC 55, 
was also observed to monitor contamination from background 
galaxies and foreground stars. Field 2 was selected to sample an area 
that is closer to the center of NGC 247 than Field 1, as an 
assessment of the Field 1 data, which were recorded one year before the Field 2 data, 
indicated that the density of NGC 247 stars in Field 1 is very small 
when compared with foreground and background objects (\S 5). 

	The GMOS data were recorded over the course of three nights during two observing 
seasons. The control field was observed through $r'$ and 
$i'$ filters during the night of October 21, 2003. Six 450 second 
integrations were recorded filter$^{-1}$, so that the 
total exposure time is 2700 sec filter$^{-1}$. Field 1 was observed on the night of 
November 23, 2003. Six 450 sec exposures were recorded 
through the $r'$ filter; $i'$ observations of this field were also scheduled, 
but these were not executed because of deteriorating image quality on this night. 
Field 2 was observed during the night of December 16, 2004. 
Nine 330 sec exposures were recorded through $r'$ and $i'$ filters, and the 
total exposure time for these data is 2970 seconds filter$^{-1}$. 

	The GMOS data were processed with tasks in the Gemini IRAF package. The basic 
reduction sequence consisted of (1) the normalization of the CCD gain values, (2) bias 
subtraction, (3) the division by a twilight flat-field frame, and (4) the removal of 
interference fringes from the $i'$ data. This latter step was done with a calibration 
frame that was constructed by median-combining the $i'$ NGC 247 images, which were 
recorded with a $2 \times 3$ point dither pattern to facilitate the suppression 
of stars and galaxies when the data were combined without spatial registration. The 
processed data were spatially aligned to correct for 
dither offsets, and then median combined to reject cosmic rays. 
Stars in the final Field 1 data have FWHM = 0.9 arcsec, while those in the final Field 
2 data have FWHM = 0.7 arcsec in $r'$.

\section{PHOTOMETRIC MEASUREMENTS}

	The brightnesses of individual sources were measured with the point spread 
function (PSF)-fitting program ALLSTAR (Stetson \& Harris 1988). The stellar 
co-ordinates, initial brightnesses, and PSFs used in ALLSTAR 
were obtained from routines in the DAOPHOT (Stetson 1987) package. 
A single PSF was constructed for each field $+$ filter combination in the CFHTIR and GMOS 
datasets. However, the PSF varies across the MegaPrime field, and so 
separate PSFs were constructed for each MegaCam CCD.

	The GMOS data were calibrated using 
observations of Landolt (1992) standards, which were recorded 
during the October 2003, November 2003, and December 2004 
GMOS observing runs. The MegaPrime data were calibrated using the photometric 
information that was placed in the image headers during Elixir processing, which 
was obtained from observations of Landolt (1992) standard star fields that 
are recorded during each MegaPrime observing run. 
The calibration of the GMOS and MegaPrime data was checked by comparing 
the brightnesess of stars common to both datasets. The mean differences between the 
two datasets are $\Delta i' = 0.02 \pm 0.02$, and $\Delta (r'-i') = 0.00 \pm 0.02$, 
where the uncertainties are standard deviations about the mean. Thus, 
the $r'$ and $i'$ calibrations of the CFHT and Gemini data are consistent.

	The photometric calibration of the CFHTIR data is based on observations of 
standard stars from Hawarden et al. (2001), which were recorded during the course of 
the three night November 2002 observing run. The instrumental $K'$ measurements were 
transformed into $K$ magnitudes. The calibration was checked against 
measurements in the 2MASS Point Source Catalogue (Cutri et al. 2003), and there is 
excellent agreement between the CFHTIR and 2MASS magnitudes and colors. In particular, the 
differences in brightness and color for sources with $K 
< 15.8$ in the three CFHTIR fields, in the sense CFHTIR -- 
2MASS, are $\Delta K = 0.04 \pm 0.13$, $\Delta (H-K) = -0.05 \pm 0.06$, and $\Delta 
(J-K) = -0.03 \pm 0.08$. The uncertainties are standard deviations about the mean. 

	Comparisons were also made with the sample of stars having $K < 17.2$ near the 
center of NGC 247 that were observed by Davidge \& Courteau (2002) with the CFHT 
adaptive optics system and KIR imager. The mean differences between the 
two datasets, in the sense CFHTIR -- Davidge \& Courteau, are $\Delta K = -0.05 \pm 0.15$ 
and $\Delta (J-K) = 0.01 \pm 0.11$. The 
uncertainties are standard deviations about the mean.

	The good agreement between the $K-$band brightnesses 
measured in this study and by Davidge \& Courteau (2002) is significant, given that 
the stars are located within 15 arcsec of the nucleus of NGC 247. 
The $K-$band images used by Davidge \& Courteau (2002) have FWHM = 0.5 arcsec, compared 
with FWHM = 1.0 arcsec for the CFHTIR data. That the $K-$band 
measurements are in good agreement, despite the difference in 
angular resolution, suggests that blending is likely not an issue among 
stars with $K < 17.2$ in this field. This conclusion is confirmed by 
artificial star experiments, which are discussed in \S 4.

\section{THE STAR-FORMING DISK}

	Carignan \& Puche (1990) found that the HI surface density in NGC 247
is flat within the central 8 kpc of the galaxy, and then drops, such that 
at a distance of 12 kpc it is one half of the value in the 
central regions. For the present study, the disk of NGC 247 has 
been divided into two regions based on the point at which the HI surface density 
drops to one half of its central value: the inner disk (R $< 12$ kpc), and the 
outer disk (R $> 12$ kpc). The stellar content within the inner 
disk is examined in this section, while the stellar content of the outer disk 
is investigated in \S 5.

\subsection{The Visible and Near-Infrared CMDs}

	The $(r' g'-r')$ and $(i', r'-i')$ CMDs of stars in various annuli 
are shown in Figure 2. The $(r', g'-r')$ CMDs are dominated by a blue 
plume that is made up of main sequence stars and blue supergiants (BSGs), 
and a red plume, made up of red supergiants (RSGs). 
While the blue plume is seen in the CMDs at all radii, the 
RSG sequence is not seen in the $(r', g'-r')$ CMD of the innermost annulus. 
The CMDs tend to go fainter with increasing galactocentric radius, 
due to the decrease in stellar density with increasing galactocentric radius. 

	The RSG and main sequence plumes are better defined on the $(i', 
r'-i')$ CMDs than on the $(r', g'-r')$ CMDs. This is likely due to the diminished 
sensitivity to reddening towards longer wavelengths. Non-uniform 
extinction is not unexpected in NGC 247, as the disk is inclined to the line of 
sight at an angle of 75.4$^{o}$ (Carignan 1985), thereby increasing the chance that 
the line of sight will pass through multiple dust clouds. 

	A population of objects that are redder than the RSG sequence appears when $i' > 
22.5$ in the $(i,' r'-i')$ CMDs. Inspection of the $9 - 12$ kpc interval CMD, where the 
density of objects is low enough to allow individual sequences to be identified, shows 
that the red stars form a tongue that is distinct from the RSG sequence. These objects are 
not simply faint RSGs; rather, the upper envelope of these red stars occurs 
near M$_{i'} \sim -5.5$, and thus coincides with the 
predicted peak brightness of the AGB (e.g. Girardi et al. 2004). 
Hence, the reddest stars at this brightness are likely evolving on the AGB.

	The $(K, H-K)$ and $(K, J-K)$ CMDs of the CFHTIR fields are compared in 
Figure 3. The near-infrared CMDs of NGC 247 are qualitatively similar to those 
of other late-type galaxies (e.g. Davidge 2005), in that there is a single 
dominant RSG sequence with $J-K \sim 1$. The bright main 
sequence that is a prominent feature in the MegaPrime CMDs is not seen in 
the near-infrared CMDs, as these hot stars are relatively faint at these wavelengths.

	The peak brightness of RSGs in the Center and North field occurs near $K = 
16.5$, while the peak of the RSG sequence in the southern field is slightly 
fainter. The population of objects with $K < 16$ and relatively blue colors 
likely consists of foreground stars and star clusters. 
Indeed, clusters of stars with diameters that are significantly smaller than 
the seeing disk, which corresponds to a linear size of $\sim 18$ parsecs with the 
adopted distance for NGC 247, will appear as a single unresolved object; the 
identification of these objects with the current data is problematic, and so some 
of the brightest objects may be such clusters. The 
presence of star clusters among the brightest objects is consistent with the tendency for
objects with $K < 16$ to be concentrated in the central field. 

	Artificial star experiments were run to assess sample completeness 
and estimate the random uncertainties in the photometric measurements. 
The results of these experiments can also be used to assess the frequency of 
blends, in which two or more stars fall within the same angular 
resolution element, and hence appear as an object that is brighter than 
either of the progenitors. The photometric properties of the artificial stars 
were measured in the same way as those of actual stars. For the purpose of computing 
completeness, an object was considered to be detected if it was recovered in two filters: 
$H$ and $K$ for the CFHTIR data, and $r'$ and $i'$ or $g'$ and $r'$ for the MegaPrime 
data.

	The effects of stellar density on the photometric measurements can be demonstrated
by considering artificial star experiments that were run on the CFHTIR Center field, 
where the stellar density is highest, and the CFHTIR North field, which is near the outer 
edge of the detected HI disk. Completeness curves in the $K-$band 
for artificial stars with colors that fall along the RSG sequence 
are shown in the top panel of Figure 4 for the Center and North fields. The 
completeness curves for both fields are similar, with incompleteness becoming significant 
when $K > 17.5$. While sample incompleteness is significant 
at $K = 17.5$, the photometric measurements are not prone to large systematic 
errors at this brightness, as the difference between the measured and actual 
brightness is $\Delta K = 0.01 \pm 0.02$ in the Center field, 
and $\Delta K = 0.02 \pm 0.02$ in the North field. 
The quoted uncertainties are the formal errors in the mean.

	If the degree of crowding increases rapidly when $K > 17.5$ then some of the 
sources at $K = 17.5$ may be blends of fainter objects. 
However, the artificial star experiments suggest that the vast majority of stars 
with $K = 17.5$ in both the Center and North fields are likely not blends. 
This is demonstrated in the lower panel of Figure 4, where the differences 
between the known and measured brightnesses of artificial stars with $K = 18$ are shown. 
Blending and the tendency to detect stars that are on positive noise spikes cause the 
$\Delta K$ distribution to be skewed to positive values near the faint limit of any 
dataset. Such a trend is seen in the lower panel of Figure 4, although $\Delta K$ is 
shifted towards positive values by only $\sim 0.1$ magnitudes on average. Obvious 
blends, in which the artificial star brightness is shifted to brighter values by a 
few tenths of a magnitude, are abscent. These comparisons indicate that the 
photometric properties of the Center and North field are very similar, despite 
differences in the density of faint stars; thus, the faint limit 
of the CFHTIR data is set by photon statistics, rather than by source confusion. 

	The situation is different for the MegaPrime data, which sample an 
intrinsically fainter population of stars than the CFHTIR data. 
The completeness curves in $i'$ for the portions of the Megaprime data that 
correspond to the CFHTIR Center and North fields 
and the scatter in the difference between actual and recovered brightnesses when $i' = 
23.0$ are shown in Figure 5. Incompleteness sets in $\sim 0.5$ magnitudes brighter 
in the Center field than in the North field. Nevertheless, the photometric measurements 
appear not to be prone to large systematic errors at $i' \sim 22.5$, as the artificial 
star experiments indicate that $\Delta i' = 0.07 \pm 0.06$ when $i' = 22.5$ in the Center 
field, and $\Delta i' = 0.04 \pm 0.02$ at the same brightness in the North field. 
As before, the quoted uncertainties are the formal errors in the mean.

	The effects of blending are apparent in the MegaPrime data near the 
center of NGC 247 when $i' > 22.5$. This is demonstrated in the lower panel of Figure 5, 
where the distributions of $\Delta i'$ values for artificial 
stars with $i' = 23$ in both the Center and North fields are compared. 
There is a large skew to positive $\Delta i'$ values in the Center field 
measurements, with some artificial stars being recovered that are 1.5 
magnitudes brighter than their actual values. These objects have almost certainly been 
blended with fainter stars. For comparison, the $\Delta i'$ distribution in the North 
field is much narrower than in the Center field, with 
the majority of recovered stars having $\Delta i' \sim 0$. The differences between the 
$\Delta i'$ distributions of the Center and North fields, coupled with the differences 
in the completeness limits shown in the top panel of Figure 5, are not unexpected if the 
faint limit of the MegaPrime data is set by crowding, rather than photon noise. 

\subsection{Comparisons with models, and the radial distribution of young stars.}

	The $(M_{r'}, g'-r')$ and $(M_{i'}, r'-i')$ CMDs of two annuli in the HI disk of 
NGC 247 are shown in Figure 6. A distance modulus of 27.9 has been adopted, based on the 
brightness of the RGB-tip measured in \S 5. A total extinction A$_B = 0.46$ magnitudes, 
which was computed for NGC 247 by Pierce \& Tully (1992) and accounts for internal 
extinction using the statistical formulation of Tully \& Fouque (1985), has also been 
assumed. The extinction in each filter was calculated using the entries in Table 6 of 
Schlegel, Finkbeiner, \& Davis (1998), which in turn are based on R$_V = 3.1$ 
extinction laws from Cardelli, Clayton, \& Mathis (1989) and O'Donnell (1994). 

	Also shown in Figure 6 are Z=0.008 isochrones from Girardi et al. (2004) with 
log(t$_{yr}) = 6.8$, 7.2, and 7.6. There are no spectroscopic abundance 
determinations for NGC 247, and so the metallicity for the models was 
selected based on the integrated brightness of the galaxy, which is comparable to that of 
M33. The HII regions in M33 have oxygen abundances that are on average $\sim 40\%$ the 
solar value (Zaritsky, Elston, \& Hill 1989); therefore, Z=0.008 models were selected 
for this study. As it turns out, the isochrones with this metallicity are roughly 
consistent with some of the sequences on the $(i', r'-i')$ CMDs. In 
particular, the Girardi et al. (2004) models predict that the $r'-i'$ color of RSGs at a 
given age is sensitive to metallicity, and the Z=0.008 models shown in Figure 6 give a 
better match to the color of the RSG sequence in NGC 247 than either the Z=0.019 or 
Z=0.004 models. That the RSG sequences in the 3 -- 6 and 9 -- 12 kpc intervals have 
comparable $r'-i'$ colors indicates that the RSGs in these two different regions have 
metallicities that agree to within $\sim 0.1$ dex.

	The good match in color between the isochrones and the blue edge of the 
main sequence in the $(r', g'-r')$ CMD is worth noting, as it suggests that the adopted 
reddening is reasonable. There is also a smattering of blue stars above the 
log(t$_{yr}) = 6.8$ isochrone, indicating that stars more massive than 30 M$_{\odot}$ 
have formed in NGC 247 during the past $\sim 6$ Myr. In \S 6, the number of bright main 
sequence stars is used to estimate the recent star formation rate (SFR) in NGC 247.

	The similar $r'-i'$ colors of the RSG sequence at various galactocentric 
distances suggests that a gradient in stellar metallicity is not present in NGC 247. This 
can be tested further by examining the relative numbers of bright blue and red stars. 
It has been established empirically that the ratio of red to blue 
supergiants varies with metallicity, in the sense that the ratio of RSGs to BSGs 
increases as metallicity drops (e.g. Humphreys \& McElroy 1984; Langer \& Maeder 
1995; Eggenberger, Meynet, \& Maeder 2002). 
The ratio of RSGs to BSGs increases in this manner because  
stars spend more of their He-burning lifetimes as RSGs than as BSGs. Consequently, a 
similar trend should also be seen in the ratio of RSGs to main sequence stars; 
if there is a radial metallicity gradient among the youngest stars 
in NGC 247 then the number of RSGs with respect 
to main sequence stars would be expected to increase with increasing radius. 

	The RSG plume in the $(M_{i'}, r'-i')$ CMDs appears 
to become more prominent with respect to the blue stellar content as 
galactocentric distance increases in NGC 247, hinting 
that there might be a gradient in the ratio of RSGs to bluer stars. 
However, this result is not statistically significant. To demonstrate this, the 
numbers of stars with M$_{i'}$ between --6.4 and --7.4 in two color intervals 
that sample blue stars and RSGs were computed. 
The blue stars were assumed to have $(r'-i')_0 < 0.1$, while RSGs were assumed to 
have $(r'-i')_0$ between 0.1 and 0.9. The resulting n$_{MS}$/n$_{RSG}$ ratios are shown 
as functions of R$_{GC}$ in the upper panel of Figure 7, and it can be seen 
that n$_{MS}$/n$_{RSG}$ does not vary with radius in NGC 247. 
Thus, the ratio of blue to red stars is consistent with there being no 
radial metallicity gradient among young stars in the disk of NGC 247.

	The luminosity function (LF) of main sequence stars is a basic diagnostic of the 
recent star-forming history of galaxies, as large changes in the SFR 
will cause discontinuities in the LF. The $r'$ LFs of stars with $(g'-r')_0 < 0.2$, 
in radial intervals with de-projected widths of 3 kpc are compared in Figure 
8. The LFs have been scaled so that they have the same 
number of stars with M$_{r'}$ between --6.1 and --4.6 to facilitate comparisons 
between the various radial intervals. To provide a benchmark for comparisons, 
a power-law was fit with the method of least squares to the M$_{r'}$ LF of all main 
sequence stars in the HI disk of NGC 247 having M$_{r'}$ between --7.1 and --4.6, and the 
result is the dashed line that is shown in each panel of Figure 8. The power-law 
exponent computed from these data is $x = 0.52 \pm 0.02$, which is slightly flatter 
than what is seen in other galaxies (e.g. Freedman 1985) \footnote[1]{Freedman (1985) 
measured the power-law exponents of main sequence stars in $U, B,$ and $V$, 
whereas the power-law measured for NGC 247 is based on $r'$ data. However, the entries 
in Table 2 of Freedman (1985) show modest -- if any -- wavelength dependence, 
suggesting that comparisons can be made with the $r'$ data. Indeed, the M$_{i'}$ LF of 
main sequence stars in NGC 247 can be fit by a power law with exponent $0.54 \pm 0.03$.} 

	The fitted power-law matches the majority of the LFs in Figure 8. The only 
major disagreement is among stars with M$_{r'} < -7.6$ in the innermost 
annulus, suggesting that there is a relative deficiency of stars that formed 
within the past $\sim 10$ Myr in this part of NGC 247. A deficiency of young stars 
is also seen in the central regions of other nearby late-type galaxies 
(e.g. McLean \& Liu 1996; Davidge 1998; Butler et al. 2004). The absence of obvious 
discontinuities in the main sequence LFs at fainter magnitudes suggests that 
star formation occured throughout the HI disk of NGC 247 during the past $\sim 40$ Myr. 
Also, the way that the star formation rate varies with time 
likely has not changed with radius -- if it had, then the LFs in Figure 8 would have 
different power-law exponents.

	While the LFs in Figure 8 suggest that recent star formation has 
occured throughout the HI disk during the past few tens of Myr, 
the radial distribution of young stars differs from that of old stars, 
indicating that there is an age gradient in the disk of 
NGC 247. To demonstrate this point, the radial distributions 
of (1) the number of bright main sequence stars, which are the prime tracer of 
young stellar content, and (2) local disk mass, which 
can be estimated from $K-$band surface photometry measurements if it is 
assumed that the near-infrared light is dominated by intermediate and old age 
populations, were compared. The results of this comparison are shown in the middle 
panel of Figure 7, where the number of main sequence stars with M$_{i'}$ 
between --6.4 and --7.4 per local stellar mass density, 
with the latter computed from the $K-$band brightness profile given in the 2MASS 
Extended Source Catalogue (Jarrett et al. 2000) and assuming M/L$_K = 2$ 
(Mouhcine \& Lancon 2003), is shown as a function of distance along the major axis.  

	If the relative number of main sequence stars per unit total local stellar 
mass does not vary with radius then the points in the middle panel of Figure 7 should 
form a horizontal sequence, with no systematic radial trend. 
However, it can be seen in the middle panel of Figure 7 that the relative density of 
bright main sequence stars per integrated local stellar mass climbs with increasing radius 
by two orders of magnitude within 12 kpc. This indicates that the 
young and old populations are not uniformly mixed throughout the disk of NGC 247. 
Rather, there is an age gradient, in the sense that the fractional mass contributed by 
young stars increases with increasing radius. This is qualitatively consistent 
with what has been inferred from integrated light studies of disks (e.g. Bell \& de 
Jong 2000). In \S 5 it is demonstrated that this trend likely breaks down outside of the 
detected HI disk in NGC 247.

	A radial variation in metallicity could introduce a systematic trend in 
the absence of an age gradient, as M/L$_K$ depends on metallicity. However, models from 
Mouhcine \& Lancon (2003) indicate that M/L$_K$ changes by only 0.2 dex 
when metallicity changes by $\Delta$[M/H]$=1.0$ dex at a given age 
for systems that are older than 10$^{9}$ years. In any event, it was demonstrated earlier 
that there is no evidence for a metallicity gradient in NGC 247.

	The density of star-forming material is of critical importance for triggering 
star-forming instabilities, and so the number of young stars in NGC 247 might be expected 
to scale with the amount of available gas and dust. To investigate if such a trend 
is present in NGC 247, the radial behaviour of the number of bright main 
sequence stars per projected HI mass, where the latter has been computed from 
the entries in Figure 4 of Carignan \& Puche (1990), is shown in the lower panel 
of Figure 7. The density of bright main sequence stars per unit HI mass drops 
by a factor of five with increasing radius. This trend is discussed further in \S 6.2. 

	The CMDs constructed from the CFHTIR data provide a means of checking the 
age and spatial distribution of the youngest stars. 
The $(M_K, J-K)$ CMDs are compared with Z = 0.008 log(t$_{yr}$) = 6.8 and 7.0 isochrones 
from Girardi et al. (2002) in Figure 9. There is a spray of relatively blue objects with 
M$_K < -12$ that are likely a mix of foreground stars and clusters. The 
main body of the RSG plume in Figure 9 extends to M$_K \sim -11.5$ in the Center
and North fields, which is roughly consistent 
with the approximate upper magnitude limit of RSGs in NGC 55 (Davidge 2005) and 
the central few arcsec of NGC 2403 (Davidge \& Courteau 2002). While the log(t$_{yr}$) = 
6.8 isochrone is bluer than the majority of stars, the log(t$_{yr}$) = 7.0 isochrone 
provides a better match to the color of the RSG plume, although this isochrone tends to 
have $J-K$ colors that are $\sim 0.1$ magnitudes too small. Isochrones with older ages do 
not give a better match to the RSG plume; rather, the RSG peak brightnesses simply become 
fainter. Adopting a higher metallicity also does not produce a better match to the 
mean $J-K$ color. The comparisons in Figure 9 indicate that the brightest red 
stars in these fields have ages log(t$_{yr}$) = 7.0, and thus confirm that there 
has been recent star formation throughout the HI disk of NGC 247. While there is a spray 
of stars above the peak isochrone brightness, it is likely that some of these objects 
will be photometric variables.

	Like the brightest main sequence stars, the number density of the brightest RSGs 
climbs with increasing distance in the HI disk. In the CFHTIR Center field there 
are 29 RSGs with M$_K$ between --10.5 and --11.0 in a region with an integrated 
brightness M$_K = -20.3$, and so there are $5.1 \times 10^{-9}$ RSGs M$_{\odot}^{-1}$. 
For comparison, there are 16 RSGs in the South field, where the integrated 
light is M$_K = -16.8$, and so the number of RSGs per unit stellar disk mass is then 
$7.1 \times 10^{-8}$ M$_{\odot}^{-1}$. There are 15 RSGs in the same magnitude interval 
in the North field, where the integrated light is M$_K = -13.9$, and so the number of 
RSGs in this field is $9.6 \times 10^{-7}$ M$_{\odot}^{-1}$. The relative numbers of 
RSGs per unit stellar disk mass in the North and South fields are thus much higher 
than near the center of NGC 247, in agreement with the trend defined by 
bright main sequence stars in Figure 7.

\section{THE OUTER DISK AND EXTRAPLANAR REGIONS}

\subsection{The Outer Disk}

	The outer disk is defined to have R$_{GC} > 12$ kpc, which is external to the 
region where Carignan \& Puche (1990) detected the bulk of HI in NGC 247. 
The $(r', g'-r')$ and $(i', r'-i')$ CMDs of two annuli in the outer disk are shown 
in Figure 10. Contamination from foreground stars and background galaxies is a 
much more significant concern than in the inner disk CMDs, especially 
near the bright end. To demonstrate the extent of this contamination, 
the CMDs of a portion of the MegaPrime field that does not contain NGC 247 stars and 
covers the same area on the sky as the $15 - 18$ kpc annulus are also shown in Figure 10.

	Bright stars in the outer disk of NGC 247 can be 
traced out to a projected distance of $\sim 18$ kpc with these data, although 
contamination from foreground stars and background galaxies becomes significant 
in the 15 -- 18 kpc annulus. For example, there are 721 objects with $i'$ between 22 and 
24 in the 15 -- 18 kpc annulus, whereas 354 objects are seen in this same 
brightness interval in the control area. While bright disk 
stars may be present at even larger distances, it becomes more 
difficult to distinguish between disk stars and contaminating objects using 
photometric data alone. Carignan (1985) measured the disk 
scale length of NGC 247 to be 4 arcmin, which corresponds to 4.4 kpc if 
the distance modulus is 27.9 (\S 5.2). Therefore, the most 
distance stars in the disk of NGC 247 detected here are at $\sim 4.1$ disk scale lengths. 

	The $(M_{r'}, g'-r')$ and $(M_{i'}, r'-i')$ CMDs of the outer disk 
regions are plotted in Figure 11, where a distance modulus 27.9 and A$_B = 0.46$ 
have been assumed. Also shown are Z=0.008 isochrones from Girardi et al. (2004) with ages 
log(t$_{yr}) = 6.8, 7.2$ and 7.6. The dominant population in the CMDs in both 
radial intervals are AGB stars, which are most obvious in the $(M_{i'}, 
r'-i')$ CMDs, and have M$_{i'} > -4.5$ and $r'-i' > 0.5$. 
Despite the noise introduced by foreground stars and background galaxies, 
a main sequence and a RSG plume can be seen in the 12 -- 15 kpc CMDs. 
The main sequence turn-off in the 12 -- 15 kpc region 
corresponds to an age log(t$_{yr}) \sim 7.2$. A corresponding population of blue main 
sequence stars is not seen in the $15 - 18$ kpc 
CMDs, and the youngest stars in this annulus have an age log(t$_{yr}) 
> 7.6$. Similarly, while a RSG sequence can be seen in the $12 - 15$ kpc 
$(i', r'-i')$ CMD, this feature is much more diffuse in the $15 - 18$ kpc annulus. 

	These data indicate that the stellar content in the inner and outer disk differ, 
in that the very bright, young stars that dominate in the inner disk 
are abscent at larger radii. Moreover, the age of the youngest 
main sequence stars increases towards larger radii in this portion of NGC 247. 
That the main sequence turn-off in the outermost 
regions of the disk probed by these data points to an age in excess of 40 Myr indicates 
that the trend for young stars to contribute progressively larger fractions to the total 
stellar disk mass towards larger radii, which was discussed in \S 4, breaks down in the 
outer disk. 

\subsection{The Extraplanar Regions}

	GMOS Fields 1 and 2 sample regions of NGC 247 that are well off of the disk 
plane. Indeed, the center of GMOS Field 2 has a projected galactocentric 
distance of 12 kpc. This is along the minor axis and, given the inclination 
of NGC 247, GMOS Field 2 then intersects the plane of the disk at a galactocentric 
distance of $\sim 75$ kpc. Consequently, any stars in the GMOS fields that 
belong to NGC 247 almost certainly do not lie along the plane of the disk.

	It can be anticipated from the CMDs in Figure 10 that foreground and 
background objects will be a dominant presence in the GMOS data. While the identification 
of individual foreground stars from photometric data alone is largely problematic, 
contamination from background galaxies can be reduced by culling obviously extended 
objects from the target lists. For the GMOS data this was done using 
the {\it daophot} `sharp' parameter, which measures the compactness of an object. 
The threshold for distinguishing between stars and extended objects was 
determined from artificial star experiments, in which the scatter in the sharp 
parameter of recovered objects was computed. It becomes more difficult to distinguish 
between stars and galaxies at progressively fainter magnitudes, due not only to the 
increased uncertainties in the sharp measurements, but also because 
faint galaxies tend to be more distant, and hence have smaller angular sizes.
Consequently, bright, highly extended objects are most effectively 
rejected with this procedure.

	Even after removing extended objects, the number 
of stars that belong to NGC 247 in GMOS Field 1 is too small to 
be measured with these data. This is demonstrated in Figure 12, where the $r'$ LFs of 
sources in GMOS Field 1 and the control field are compared. 
There is good agreement between the two LFs, indicating that the density of 
sources with comparable magnitudes in the two fields are similar; if GMOS 
Field 1 contains stars belonging to NGC 247, then they occur in numbers that 
are swamped by foreground and background objects. While GMOS Field 1 thus 
contributes little information about the extraplanar stellar content of NGC 247, the 
comparison in Figure 12 is still of interest, as the good agreement between the 
LFs suggests that the foreground star and background galaxy counts in the control 
field are representative of those near NGC 247.

	The situation is very different in GMOS Field 2, where there is an obvious 
population of objects that belong to NGC 247. This is demonstrated in Figure 13, 
where the $(i', r'-i')$ CMDs of GMOS Field 2 
and the control field are compared. There is a population of 
objects in GMOS Field 2 that have $(r'-i') < 0.1$ and $i' > 24$, that is not seen in 
the control field. Indeed, there are 571 stars in GMOS Field 2 with $i'$ between 24 and 
25.5 and $r'-i'$ between 0 and 1, while there are 363 objects in the corresponding portion 
of the control field CMD. 

	The NGC 247 stars in GMOS Field 2 occupy a much narrower range of colors 
than those of foreground stars and background galaxies. 
This is demonstrated in the lower left hand panel of Figure 13, where 
the histogram distribution of the difference in the number of 
objects with $i'$ between 24.75 and 25.25 in GMOS Field 2 and the control field is shown 
per 0.2 magnitude interval in $r'-i'$. This comparison indicates that there is 
an excess number of objects in GMOS Field 2 that have $r'-i'$ between 0.3 and 0.7; this 
range of colors is consistent with an RGB population.
The spatial distribution of objects in this color range is such that 
they are concentrated on the side of GMOS Field 2 that is closest to NGC 247, and so 
the statistical significance of objects with this range of color and brightness is 
lower in the half of the field that is furthest from NGC 247. Consequently, all 
subsequent analysis is based on the half of GMOS Field 2 that is closest to 
the main body of the galaxy.

	The CMDs in Figure 13 also differ near the faint end. A 
sequence of objects with $i' >$ 26 that extends to $r'-i' > 1$ is seen in the control 
field CMD; these faint objects likely are moderate to high redshift galaxies. However, 
only a modest corresponding sequence, that does not extend to red colors, is seen in the 
GMOS Field 2 CMD. This difference likely occurs because the two CMDs do not have the 
same faint limits. While the two fields were observed with roughly the same total 
exposure times, the control field data was recorded during better seeing conditions than 
the GMOS Field 2 data (FWHM $= 0.6$ arcsec in $r'$ and $i'$ for the control field, versus 
0.7 arcsec for GMOS Field 2), and so go deeper. Consequently, the control field data 
sample redder objects when $i' > 26$ than the GMOS Field 2 data.

	To facilitate the analysis of the GMOS Field 2 data, artificial star experiments 
were run to assess completeness and measure the errors in the photometric measurements. 
The artificial stars were assigned colors that are typical of RGB stars in NGC 247, 
and were considered to be recovered only if they were detected in both $r'$ and $i'$. 
The completeness curve in $i'$ and the dispersion in the brightnesess 
of recovered stars with $i' = 25$ are shown in Figure 14. Incompleteness becomes 
significant when $i' > 25$, while the distribution in the difference between the actual 
and measured brightnesses of artificial stars with $i' = 25$ has a dispersion of $\pm 
0.1$ magnitude. The modest dispersion in the difference between the actual and measured 
brightnesses indicates that blending is not an issue in GMOS Field 2; 
this is not unexpected given the relatively low density of sources in this field.

	The $i'$ LFs of sources in GMOS Field 2 and the control field are compared in 
Figure 15. To maximize the contrast with respect to contaminating sources, 
only those objects with $r'-i'$ between 0.3 and 0.7 were used to construct the LFs.
The comparison in Figure 15 indicates that there is an excess 
number of stars in NGC 247 with $r'-i'$ between 0.3 and 0.7 over a 
large range of brightnesses, with a marked discontinuity occuring near $i' = 24.5$. 
The size of the discontinuity is consistent with that expected between 
populations that are dominated by RGB stars ($i' > 24.5$) and AGB stars ($i' < 24.5$). 
This is demonstrated in the lower panel of Figure 15, where a power-law with exponent $x = 
0.22$, which is the exponent that was derived from the $i'$ LF of RGB stars in NGC 55 
(Davidge 2005), has been fit to the control field-corrected LF entries with $i' > 24.5$. 
The dashed line shows this same power-law relation, but shifted down assuming that AGB 
evolution is $\sim 4\times$ faster than on the RGB. The AGB sequence computed 
in this manner passes through the points with $i' < 24.5$ in the 
background-subtracted LF. Therefore, $i' = 24.5 \pm 0.1$ is identified as the 
brightness of the RGB-tip in NGC 247. 

	Stars in the outer regions of late-type galaxies tend to have [Fe/H] $\leq -1$ 
(e.g. Davidge 2003, 2005; Tiede et al. 2004; Brooks, Wilson, \& Harris 2004; Galleti, 
Bellazzini, \& Ferraro 2004), and this is also almost certainly the case for NGC 247. If 
the majority of RGB-tip stars are very metal-poor -- an assumption that is consistent 
with the mean color of these objects (see below) -- then they will have M$_{i'} \sim -3.4$ 
(Girardi et al. 2004). Hence, if a low metallicity is assumed then the distance modulus of 
NGC 247 is $27.9 \pm 0.1$.

	The correction to the distance modulus for extinction is likely to be negligible. 
Given that Field 2 is well off of the disk plane, 
internal extinction from the disk is expected to be modest. The dominant 
source of extinction is then in the foreground, where A$_B = 0.08$ magnitudes 
(Schlegel et al. 1998). This corresponds to only A$_{i'} = 0.04$ magnitudes.

	In an effort to construct a CMD of GMOS Field 2 that is 
free of foreground and background sources, objects in the 
GMOS Field 2 and the control field that are located within $\pm 0.1$ magnitude of each 
other on the CMD were paired. Objects in the control field were only paired 
with one object in GMOS Field 2, and those objects in GMOS Field 2 that were matched with 
an object in the control field were subtracted from the CMD. The $\pm 0.1$ magnitude 
matching distance was chosen as it corresponds roughly to the random uncertainties in the 
photometry near the faint end (Figure 14); however, experiments show that the 
culled CMDs are not sensitive to the pairing distance.

	The CMD of the objects that remain after this statistical culling procedure 
is shown in the left hand panel of Figure 16, 
along with log(t$_{yr}$)=10 isochrones from Girardi et al. (2004). 
It appears that the RGB stars in GMOS Field 2 have a broad range of 
metallicities, spanning from Z $\leq$ 0.0001 to at least Z = 0.004. In fact, there are 
objects to the right of the Z = 0.004 sequence, hinting that a relatively metal-rich 
population is present.

	There is a modest spray of objects above the RGB-tip 
in the left hand panel of Figure 16 that is due to stars 
evolving on the AGB; this population is also seen above the RGB-tip in Figure 15. 
A comparison with the isochrones in Figure 16 indicates that some of these AGB stars 
are simply the brightest members of an old population. However, there is a 
population of objects with $r'-i' > 1$ and $i' < 24$ in the 
statistically corrected CMD that are too bright to belong to an old population. 
In fact, if the stars with $i' < 24$ and $r'-i' > 1$ 
in the left hand panel of Figure 16 are evolving near the 
AGB-tip then they have an age log(t$_{yr}) \sim 9.5$; thus, an 
intermediate age population may be present in the extraplanar regions of NGC 247.
It should be emphasized that only a handful of objects are 
found in this region of the control field CMD, and the star count 
model described by Robin et al. (2003) predicts that no more than one 
of the objects with $i'$ between 22 and 24 and $r'-i'$ between 1 and 1.5 in the control 
field CMD is likely a Galactic star. Therefore, any object in this portion of the CMD is 
either a background galaxy or an AGB star in NGC 247. 

\section{SUMMARY \& DISCUSSION}

	Images that sample the visible and near-infrared 
wavelength regions have been used to investigate the stellar content of the spiral 
galaxy NGC 247. The data were recorded with three instruments: MegaPrime and CFHTIR, 
both of which are on the CFHT, and GMOS, on Gemini South. With the exception of gaps 
between the detector elements, the MegaPrime data sample the entire disk of NGC 247, and 
provide the first complete census of bright main sequence stars 
and supergiants in this galaxy. While not covering the entire disk, the 
CFHTIR data allow the properties of RSGs in three spatially distinct regions to 
be investigated, and provide an independent means of checking some 
of the trends found at shorter wavelengths. Finally, the GMOS data have been used to 
investigate the stellar content of the extraplanar regions. The GMOS data 
reveal a relatively broad metallicity distribution among RGB stars and a population 
of bright extraplanar objects, which likely are intermediate age AGB stars.
The distance to NGC 247 is also estimated from the brightness of the RGB-tip. 

\subsection{The Star-Formation Rate in the Disk Of NGC 247}

	A population of bright main sequence stars have been detected throughout the disk 
of NGC 247, and these can be used to estimate the recent SFR in a direct manner. 
A modest SFR might be expected given the small number of bright main 
sequence stars, and this expectation is consistent with SFRs computed from other 
estimators. Indeed, the SFR computed from the IRAS 60 and 
100$\mu$m fluxes discussed by Rice et al. (1988) is 0.01 M$_{\odot}$ year$^{-1}$; 
the low FIR flux of NGC 247 is consistent with a 
modest X-ray flux (e.g. Read \& Ponman 2001). Condon (1987) 
measured the 1.49 Ghz flux from NGC 247, which also can be used to compute the SFR. Using 
the calibration given by Condon (1992) the SFR for NGC 247 is 0.001 
M$_{\odot}$ year$^{-1}$. Finally, based on the integrated H$\alpha$ brightness, 
Ferguson et al. (1996) conclude that the SFR in NGC 247 is 0.09 M$_{\odot}$year$^{-1}$.

	To estimate the SFR from the Megaprime data, the number of stars above the 
log(t$_{yr}) = 7.2$ main sequence turn-off were counted. After correcting 
statistically for foreground and background sources, 
there are 120 objects that fall above the log(t$_{yr}) = 7.2$ isochrone in the 
$(r', g'-r')$ CMD. If all of these are assumed to be actual stars in NGC 247 then 
they account for $\sim 4800$ M$_{\odot}$. If the IMF in NGC 247 is assumed to follow a 
Miller-Scalo (Miller \& Scalo 1979) IMF, with an assumed upper mass of 100 M$_{\odot}$ 
and a lower mass of 0.1 M$_{\odot}$, then the stars above the log(t$_{yr}) = 7.2$ 
isochrone account for $\sim 0.3\%$ of the total mass of stars that ultimately will form. 
The bright main sequence stars are thus signposts of $\sim 1.7 \times 10^6$ M$_{\odot}$ 
of star formation within the past 16 Myr, and the SFR is then $\sim 0.1$ M$_{\odot}$ 
year$^{-1}$. Similar SFRs are computed using different age thresholds; for example, a SFR 
of 0.16 M$_{\odot}$ year$^{-1}$ is computed based on source counts above the 
log(t$_{yr}) = 6.8$ isochrone. While these SFRs are 
upper limits, as some of the bright stars may be clusters, 
it is encouraging that the SFR computed from the 
number of bright main sequence stars is in excellent agreement with the SFR measured by 
Ferguson et al. (1996). This agreement is significant, as both techniques are based on 
a common statistic, which is the number of bright main sequence stars. 

	While the SFR estimates based on bright stars counts, FIR flux, the strength of 
the radio continuum, and H$\alpha$ flux span two orders of magnitude, they 
all indicate that the current SFR in NGC 247 is modest. There are hints that the SFR may 
have been higher roughly 0.1 Gyr in the past, at least in the inner regions of the 
galaxy. The nucleus of NGC 247 has near-infrared colors that are similar to those of M33 
(Davidge \& Courteau 2002), suggesting similar ages. The SED of M33 indicates that 
the last star-forming activity in the nucleus of M33 occured 
70 - 75 Myr in the past (Gordon et al. 1999), and so this is likely also the age 
of the nucleus of NGC 247. However, the nucleus may not be an indicator of 
larger scale star-forming activity, as the stellar content of the inner disk may be 
de-coupled from that of the nucleus. Indeed, there is a central deficiency of 
stars with an age $< 0.5$ Gyr (Stephens \& Frogel 2002) in the inner disk of M33, even 
though stars evidently formed in the nucleus during the past 0.1 Gyr (Gordon et al. 
1999). Consequently, it is possible that any star-forming activity near the center of NGC 
247 0.1 Gyr in the past may have been confined just to the nucleus.

\subsection{The Radial Stellar Content of the Disk}

	Bell \& de Jong (2000) used surface photometry to investigate the radial 
stellar contents of spiral galaxies, and concluded that these systems contain population 
gradients, in the sense that the central regions tend to be older and more metal-rich 
than the outer regions. Studies of the resolved stellar 
contents of nearby galaxies provide a direct means of checking this result. 
Indeed, one of the key results of the current study is that the stellar content in the 
disk of NGC 247 varies with galactocentric radius, in the sense that the fractional 
contribution of young stars to local stellar mass increases with 
increasing radius in the inner disk. 

	While the trend for mean age to vary with radius in the inner disk of NGC 247 is 
consistent with what was seen in the sample of galaxies studied by Bell \& de Jong (2000), 
there is no evidence for an accompanying gradient in stellar 
metallicity. Indeed, both the $r'-i'$ color of the RSG sequence 
and the ratio of bright blue stars to RSGs are metallicity indicators, and neither 
varies with radius in the inner disk. This result could be checked further by 
obtaining deep high angular resolution images of the disk, which could be used to 
measure the colors of individual RGB stars, and determine if mean RGB color changes 
with radius. 

	An age gradient will affect the integrated colors of NGC 247 -- is there evidence 
for a radial color gradient in the surface photometry 
of NGC 247? There are two datasets available to address 
this question. Data in the 2MASS Extended Source Catalogue (Jarrett 
et al. 2000), which does not extend beyond the detected HI disk, indicates that $J-K$ 
becomes bluer with radius in NGC 247, as expected if there is an age gradient. 
For comparison, the $J-F$ profile given in Figure 15 of Carignan (1985), 
which extends out to a projected distance of $\sim 16.5$ kpc, is relatively 
flat, although there is an upturn at very large radii.

	Ferguson et al. (1996) investigated the radially-averaged H$\alpha$ profile of NGC 
247, and the results in their Figure 4 indicates that the H$\alpha$ surface brightness 
drops by roughly an order of magnitude over a 10 arcmin distance from the galaxy center. 
For comparison, the 2MASS surface brightness measurements predict that 
the $K-$band surface brightness drops by $\sim 9$ magnitudes over the same 
radial interval, which corresponds to a radial drop in stellar density that is in excess 
of three orders of magnitude. Thus, the relative density of H$\alpha$ emission per unit 
stellar mass increases by $\sim 2$ orders of magnitude from the galaxy center to the outer 
regions of the disk, providing further evidence for an age gradient in NGC 247.

	The trend for mean age to decrease with increasing radius stops at the edge of 
the detected HI disk, where massive main sequence stars disappear. An older 
stellar substrate persists $\sim 6$kpc beyond the edge of the detected HI disk, suggesting 
that mean age may drop with increasing radius in the outer disk. 
While it is difficult to trace the integrated colors of nearby 
disks out to very large radii, the age trend 
seen here is reminiscent of what can be inferred for the z=0.3 edge-on disk galaxy 
that was investigated by Zibetti \& Ferguson (2004). In that galaxy the 
color along the major axis becomes bluer with increasing radius, 
as expected if mean age decreases with increasing radius, out to a point 
where the disk scale length changes. At larger radii the color then 
becomes progressively redder with increasing radius.

	The threshold for triggering star formation based on the Toomre criterion depends 
on the surface mass density of gas (e.g. Kennicutt 1989). In the context of this 
paradigm one might then ask if the overall radial variation in mean age in NGC 247 
is consistent with the relative distribution of gas and young stars. Before attempting to 
address this issue, it should be noted that star formation is closely 
tied to the molecular component of the ISM, and so the radial 
distribution of the youngest stars should be compared with that of molecular 
gas. Unfortunately, we are not aware of a published survey of 
molecular material in NGC 247, and so HI is the only available means of tracing the ISM 
of NGC 247 at present. This is significant, as HI is not necessarily a good tracer 
of molecular gas, as it tends to be more widely distributed than molecular gas 
in field spiral galaxies, although among Virgo cluster galaxies the HI and 
molecular components have similar distributions (Nishiyama, Nakai, \& Kuno 2001). 
While NGC 247 is in a low density environment, the HI distribution of NGC 247 is 
truncated, and radially truncated HI distributions are a characteristic of spiral galaxies 
in the central regions of the Virgo cluster (e.g. Warmels 1988).
Nevertheless, given the reliance on HI measurements, rather than molecular material, the 
following discussion is restricted to the investigation of qualitative trends between 
stars and gas. 

	If HI traces the ISM of NGC 247, then the surface 
density of young stars should be roughly constant across the 
disk, and the number of such young stars with respect to the underlying stellar 
body of the disk would then grow with increasing radius in the inner disk; this behaviour 
is seen in the middle panel of Figure 7. However, the HI density plummets near 
12 kpc, and so the likelihood of star formation occuring at larger radii should 
also drop if the gas density falls below the threshold needed to trigger star 
formation. If the stellar disk extends past the point at which the HI disk ends
then the ratio of young to older stars should decrease 
when R$_{GC} > 12$ kpc, as the probability of forming young stars decreases with 
increasing radius. Such an absence of young stars is in fact seen in the outer disk of 
NGC 247. Thus, the distribution of stars and gas are in qualitative agreement with the 
expectations of a threshold-driven model of star formation.

\subsection{The Extended Stellar Disk}

	Most photometric studies of edge-on disks have suggested that the stellar 
component truncates after a few disk scale lengths (e.g. van der Kruit 
\& Searle 1981; de Grijs, Kregel, \& Wesson 2001). A possible physical explanation for 
such a cut-off is that the gas density in disks drops to a level where the critical 
threshold to trigger star-forming instabilities is not breached (e.g. Martin \& 
Kennicutt 2001). However, the existence of a sharp outer boundary in stellar disks 
has been questioned by Narayan \& Jog (2003), who argue that the observed disk 
edges may simply be artifacts of photometric depth effects. 
Indeed, deep images of barred S0 - Sb galaxies obtained by Erwin, Beckman, \& Pohlen 
(2005) indicate that some have extended disks that show no evidence of truncation.

	Observations of nearby galaxies support the notion that 
extended stellar disks may be a common phenomenon. Young and/or intermediate age stars 
are seen at R$_{GC} \sim 30$ kpc along the major axis of M31 (Ferguson \& Johnson 
2001), as well as at R$_{GC} \geq 15$ kpc in NGC 2403 (Davidge 2003) and M33 
(Davidge 2003; Tiede, Sarajedini, \& Barker 2004). Bland-Hawthorn et al. 
(2005) find disk stars at a projected distance of 14.4 kpc in NGC 300, while Gallart 
et al. (2004) find evidence for an extended disk in the LMC.
The intermediate-age stars in the outer regions of NGC 2403 and M33 are distributed over 
large areas (Davidge 2003), indicating that they did not form in isolated pockets; rather, 
the star-forming activity appears to have been wide-spread.
That globular cluster systems with disk-like kinematics are found at large 
galactocentric radii around nearby spirals (e.g. Olsen et al. 2004) further 
suggests that stellar disks can extend out to very large distances, and may even be 
relatively old (Morrison et al. 2004, but see also Beasley et al. 2004). 

	In the current study, stars with ages log(t) $\sim 7.2$ are seen outside of the 
detected HI disk of NGC 247. While stars with this age are restricted to within 3 kpc 
of the edge of the detected HI disk, stars with only slightly older ages are seen out to 
18 kpc. How did these stars form given the low density of HI seen at the 
present day? One explanation is that there may be undetected clumps of HI in the outer 
disk of NGC 247. The HI observations used by Carignan \& Puche (1990) had an 
angular resolution of 45 arcsec, and so HI pockets that are markedly smaller than this may 
have gone undetected. Such HI pockets need not be very compact to form stars, as 
spiral density waves or stochastic processes may induce 
star formation in material that might otherwise fall under the threshold 
needed to induce star formation. However, it seems unlikely that the stars in the 
outer disk of NGC 247 formed as a result of fluke perturbations in a 
clumpy HI envelope, as the stars are more-or-less uniformly distributed, 
with no evidence of large-scale clumping. Moreover, if this process was at work 
in the outer disk of NGC 247 then it might be anticipated that there would be isolated 
pockets of star formation in the outer disk at the present day, and none are seen.

	Could the youngest stars in the outer disk 
have formed at smaller galacocentric radii, and then migrated to larger radii 
as a result of, for example, viscosity and/or spiral density waves (e.g. Sellwood 
\& Binney 2002)? This is unlikely as these processes will act on timescales 
of a few disk rotation periods (i.e. $\sim 1$ Gyr), and so the youngest stars 
in the outer disk should have intermediate ages. For comparison, 
the 12 - 15 kpc annulus contains main sequence turn-off 
stars with an age of $\sim 20$ Myr. 

	Galaxies do not evolve as isolated systems, and dynamical interactions 
may deposit stars in the outer disk (e.g. Ibata et al. 2003). This mechanism 
was suggested by Erwin et al. (2005) as one possible cause of extended disks. However, 
in the case of NGC 247 these interactions must have happened within the past few tens of 
millions of years to match the age of the youngest stars in the outer disk in NGC 247. 
There is no obvious companion of NGC 247 that would have been involved 
in such an interaction. 
 
	Another possibility is that the HI disk may have been larger in the past than it 
is now. While it is a matter of speculation, interactions with NGC 253, which is 
the closest large galaxy to NGC 247, may have disrupted the HI disk of NGC 247. 
The presence of intermediate age stars in an extraplanar 
envelope (\S 5) is consistent with such an interaction (see below).
However, if the HI envelope was stripped away by encounters with another galaxy then 
there might be tidally-induced structures, and 
there is no evidence for HI bridges or tails in this part of the Sculptor 
Group (Haynes \& Roberts 1979; Putman et al. 2003).

	None of the mechanisms considered here provide a clear explanation for the 
presence of stars in the outer disk of NGC 247. Deep imaging of the HI content around NGC 
247 at angular resolutions that exceed those obtained by Carignan \& Puche (1990) 
may provide additional clues into the origin of the stars in the outer disk. For example, 
the detection of localised pockets of HI may indicate sites of potential or 
low-level on-going star formation, while the detection of large scale streams 
or filaments may be signatures of a disrupted HI disk.

\subsection{The Extraplanar Regions}

	In \S 5 it is demonstrated that GMOS Field 2 likely contains bright AGB stars 
with ages log(t) $\sim 9.5$. Stars with roughly similar ages are seen in envelopes around 
other spiral galaxies. Dalcanton \& Bernstein (2002) examine a sample of bulge-less 
edge-on disks and find that they have extended envelopes that contain stars that (1) are 
moderately metal-rich, suggesting a disk origin, and (2) have ages indicating
that they formed $\sim 6 - 8$ Gyr in the past.
Brown et al. (2003) find a population of stars in the extraplanar regions of M31 
that have an age of $6 - 8$ Gyr based on the location of the main sequence turnoff (MSTO) 
in the CMD, although Ibata et al. (2004) note that the stars in this portion of M31 
may belong to a stellar stream, rather than to a uniformly distributed component.
While the Galactic thick disk appears to have an age $\geq 10$ Gyr (e.g. Liu \& Chaboyer 
2000; Bensby, Feltzing, \& Lundstrom 2003), it may contain stars that formed 
$\sim 5$ Gyr after its creation (Bensby, Feltzing, \& Lundstrom 2004). 
However, not all disk systems appear to be surrounded by intermediate-age envelopes, as 
the resolved AGB content in the extraplanar regions of NGC 55 is suggestive of an 
old age (Davidge 2005). 

	Envelopes containing intermediate age stars may result from 
interactions between a host galaxy and its companions. For example, interactions 
with a companion may heat the stars in the disk of the host galaxy, causing them to 
leave the disk plane (e. g. Quinn \& Goodman 1986; Bekki \& Freeman 2003). 
Extraplanar envelopes around the host galaxy may also result from mergers 
between satellites (e.g. Bekki \& Chiba 2001) or the stripping of material from a 
satellite (Tsuchiya, Dinescu, \& Korchagin 2003).

	There are hints that spiral galaxies in what appear to be relatively low 
density environments at the present day did not evolve in isolation. In the case 
of the Milky-Way there are a handful of younger Galactic globular clusters that are 
thought to have been accreted from companion galaxies (e.g. Searle \& Zinn 
1978), and some of the globular clusters in the Galactic halo may have 
originated in the Sagittarius dwarf (Bellazzini, Ferraro, \& Ibata 2003). 
Interactions between the Milky-Way and its companions have left debris trails, such 
as the one associated with the Sagittarius dwarf (e.g. Ibata et al. 2001; 
Martinez-Delgado et al. 2001), and possibly even from a now-defunct dwarf 
galaxy that may have been the progenitor of $\omega$ Cen (e.g. Mizutani, Chiba, 
\& Sakamoto 2003; Bekki \& Freeman 2003). The satellites that have been consumed 
by the Milky-Way may have been relatively massive. Indeed, Gilmore, Wyse, 
\& Norris (2002) suggest that the satellite involved in the 
interactions that are hypothesized to have formed the Galactic thick disk may have 
had a mass approaching a quarter of that of the Milky-Way. 

	Could the extraplanar stars in NGC 247 have formed {\it in situ}? 
The extraplanar regions around spiral galaxies may 
contain gas that was ejected from the disk, or may have been accreted from 
an outer corona, although this likely is not a major contributor to halo 
mass. For example, in the case of M31, gas of this nature in the halo may amount 
to $\sim 1\%$ of that in the HI disk (Thilker et al. 2004). Shocks from areas 
of vigorous star-forming activity in the disk may trigger star formation 
in clouds that are well off of the disk plane (e.g. Tullmann et al. 2003). 
However, the observed low density of HII regions in extraplanar environments suggests 
that the overall SFR off of the disk is low (e.g. Rossa et al. 2004). 
The wide metallicity spread in GMOS Field 2 indicates that if the stars did form 
{\it in situ}, then they likely did so from gas that originated in the disk or 
a large companion. 

	The spatial distribution of objects in GMOS Field 2 that belong to NGC 247 
will provide clues into the origin of these stars. For example, 
a sharp drop in the number counts over a small distance would be one signature of 
a tidal stream. While the detection of such an edge is 
complicated by the large number of contaminating objects, the data appear to be 
consistent with a smooth decrease in number counts across the field, rather than a 
sharp discontinuity. Deep images with higher angular resolution, such as could be 
obtained with the HST ACS, would allow relatively compact 
background galaxies, which appear as point sources in the MegaCam data, to 
be culled from the sample, thereby increasing the contrast 
between stars in NGC 247 and contaminating objects.

	We close this part of the discussion by noting that the GMOS data sample 
only a modest portion of the extraplanar regions of NGC 247, and 
a deep, wide-field survey of the outer regions of the galaxy 
would provide further insight into the nature of the brightest AGB stars. 
For example, the detection of organized structures would argue for a tidal origin 
for the intermediate age component. On the other hand, if the brightest AGB stars 
are found to be restricted to a modest area that overlaps with the GMOS field then this 
would suggest that they may belong to an extended very low surface brightness satellite of 
NGC 247. 

\subsection{The Distance Modulus of NGC 247}

	Previous distance estimates for NGC 247 have relied on indirect methods
or indicators that are based on the global properties of the galaxy. de Vaucouleurs 
(1978) computed a distance modulus of 27.0 for the Sculptor Group in general, while 
de Vaucouleurs (1979) found specific distance moduli between 26.3 and 
27.2 for NGC 247 from different tertiary distance indicators. 
More recent studies suggest a larger distance modulus for NGC 247. Karachentsev 
et al. (2003) compute a distance modulus of 28.06 based on the Tully-Fisher 
relation. They also point out that NGC 247 is likely part of the 
NGC 253 sub-group, and find a distance modulus of 28.0 for NGC 253 based on the RGB-tip. 
If an undisturbed local Hubble flow with H$_0 \sim 65$ is assumed 
then the distance modulus of NGC 247 is 27.6. 

	The distance modulus measured for NGC 247 from the brightness of the RGB-tip is 
$27.9 \pm 0.1$, and this is consistent with the distance 
modulus of NGC 253 measured by Karachentsev et al. (2003). A less 
secure distance modulus can also be computed from the peak brightness of 
RSGs, which the CFHTIR data indicates occurs near $K \sim 16.5$. The calibration of 
Rozanski \& Rowan-Robinson (1994) indicates that the brightest RSGs in a galaxy with an 
integrated brightness equal to that of NGC 247 will have M$_K \sim -11.6 \pm 0.5$, and so 
the distance modulus of NGC 247 from RSGs is $28.0 \pm 0.5$. This is consistent with the 
RGB-tip distance modulus.

\acknowledgements{It is a pleasure to thank the anonymous referee for a report that 
lead to sgnificant improvements in the manuscript.}

\parindent=0.0cm

\clearpage

\begin{table*}
\begin{center}
\begin{tabular}{llcc}
\tableline\tableline
Instrument & Field & RA & Dec \\
 & & (J2000) & (J2000) \\
\tableline
CFHTIR & North & 00:47:04.5 & --20:36:00.5 \\
 & Center & 00:47:12.1 & --20:45:04.3 \\
 & South & 00:47:12.2 & --20:51:00.8 \\
 & & & \\
GMOS & Field 1 & 00:46:28.0 & --20:48:27.0 \\
 & Field 2 & 00:46:42.3 & --20:46:10.5 \\
\tableline
\end{tabular}
\end{center}
\caption{Field Locations}
\end{table*}

\clearpage

\begin{figure}
\figurenum{1}
\epsscale{1.0}
\plotone{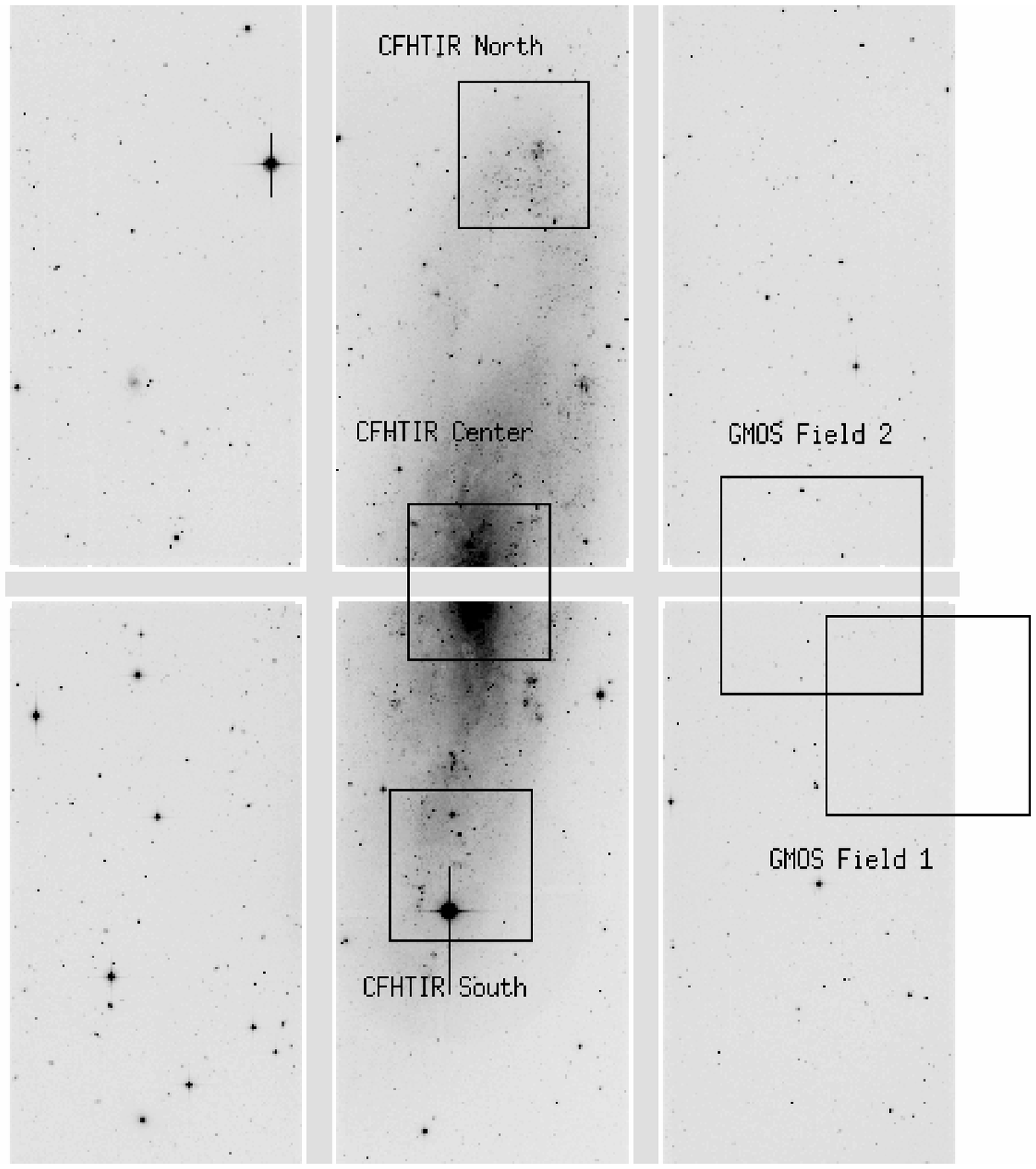}
\caption
{The six elements of the MegaCam mosaic that sample NGC 247 and its immediate 
surroundings. The data shown here were recorded through the $r'$ filter. 
North is at the top, and east is to the left. The six elements 
cover roughly $20 \times 30$ arcmin$^2$. The locations of 
the fields observed with CFHTIR and GMOS are also indicated.}
\end{figure}

\begin{figure}
\figurenum{2}
\epsscale{1.0}
\plotone{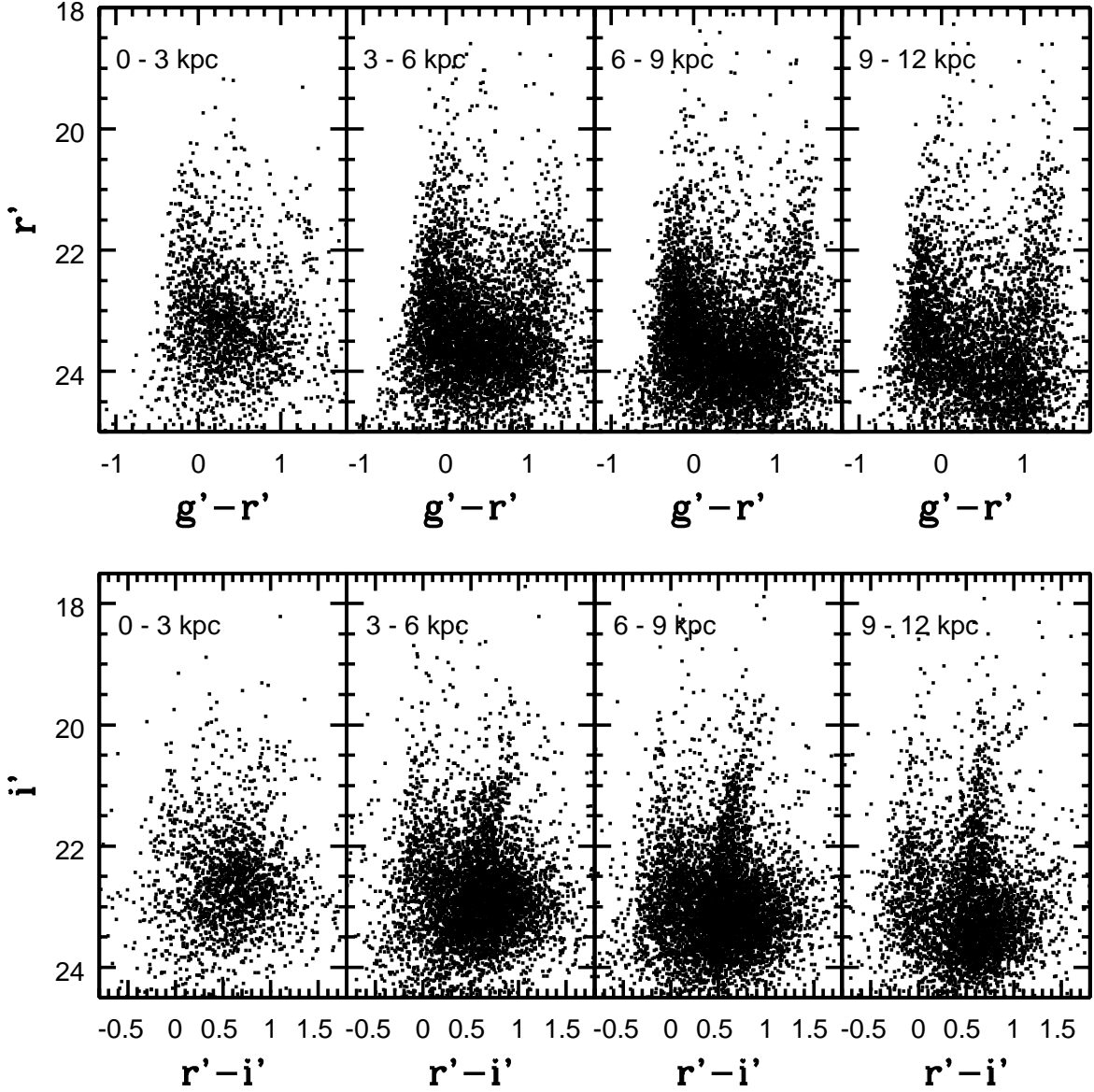}
\caption
{The $(r', g'-r')$ and $(i', r'-i')$ CMDs of stars throughout the inner disk of NGC 247. 
The distance intervals indicated in each panel are de-projected galactocentric radii, 
assuming a distance modulus of 27.9 (\S 5) and an inclination of 75.4$^{o}$ 
(Carignan 1985). The blue plume in each CMD contains bright main sequence stars and 
BSGs, while the red plume contains RSGs. The brightest AGB stars form the clump 
of objects in the $(i', r'-i')$ CMD with $i' > 22.5$ and $r'-i' > 0.5$.}
\end{figure}

\begin{figure}
\figurenum{3}
\epsscale{1.0}
\plotone{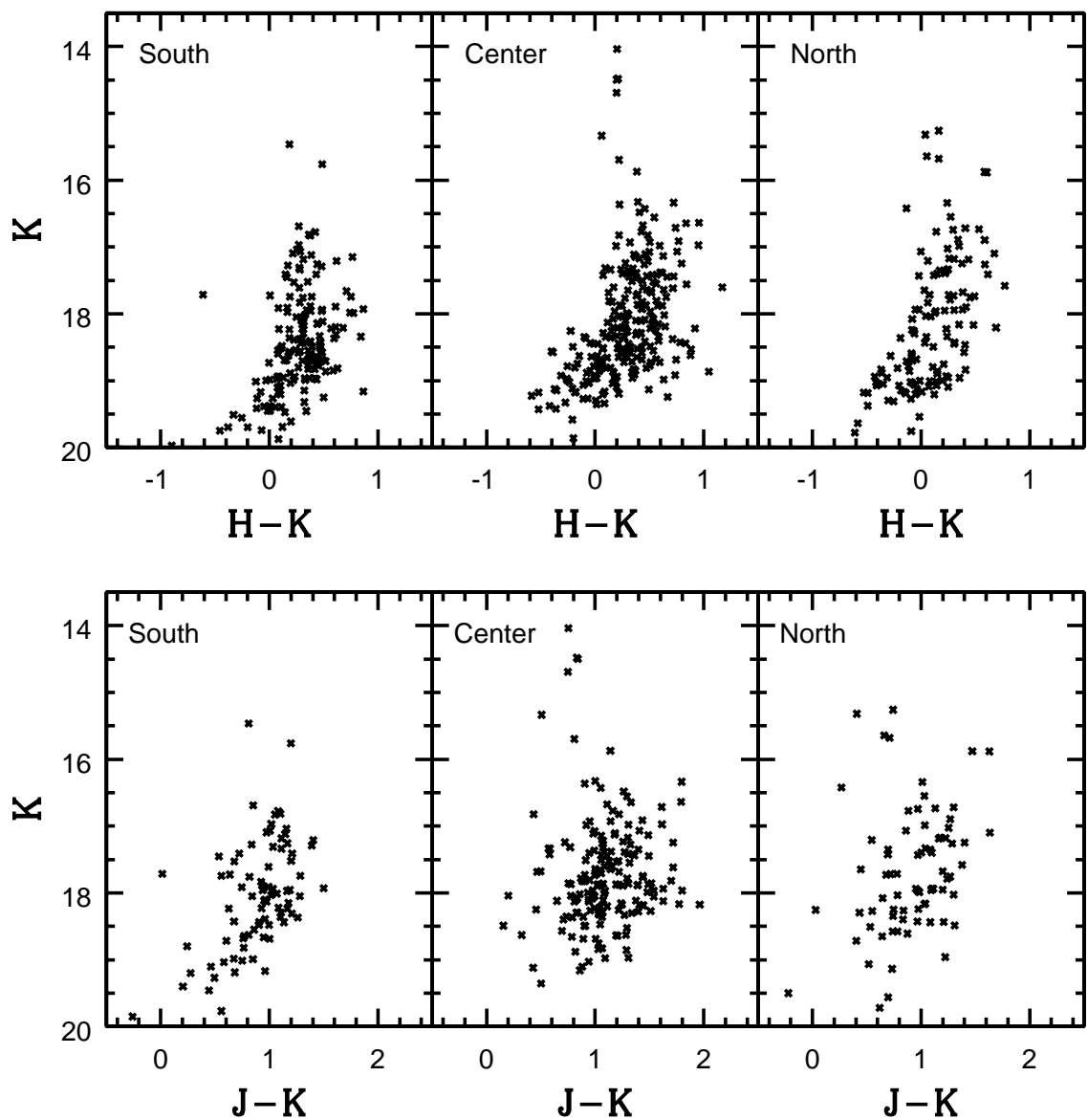}
\caption
{The $(K, H-K)$ and $(K, J-K)$ CMDs of the South, Center, and North CFHTIR fields. The 
sources with $K < 16$ are likely foreground stars and unresolved star clusters, while 
the majority of sources with $K > 16.5$ are RSGs in NGC 247.}
\end{figure}

\begin{figure}
\figurenum{4}
\epsscale{1.0}
\plotone{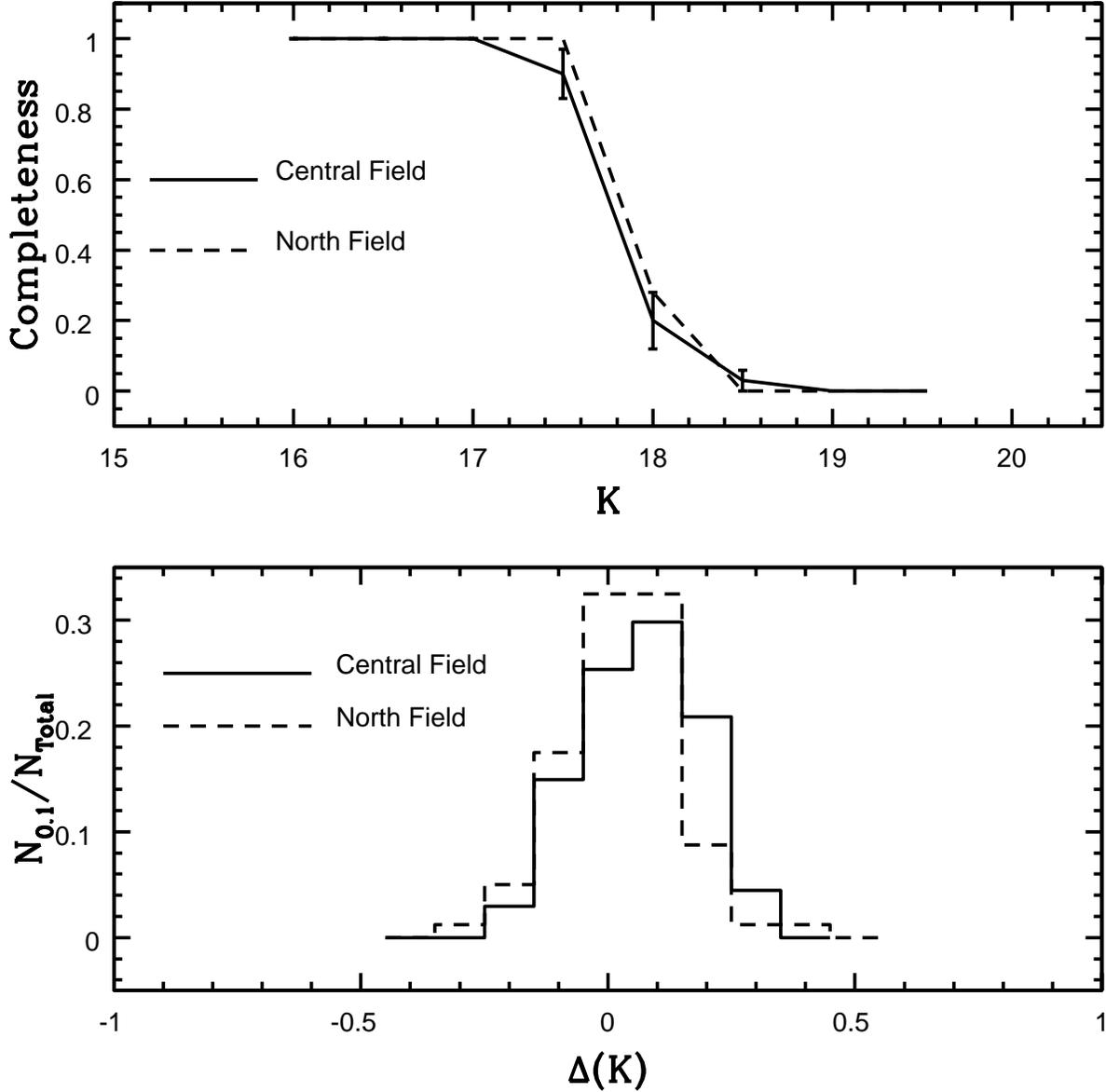}
\caption
{The results of artificial star experiments run on the CFHTIR 
observations of the Center and North fields. The top panel shows the completeness fraction 
in $K$, where completeness is the ratio of the number of recovered artificial stars 
to the number that were added. To be recovered, an artificial star had to be detected 
in both $H$ and $K$. The lower panel shows the difference between the actual and measured 
brightnesses of artificial stars, $\Delta K$, with $K = 18$, which is 0.5 magnitudes 
fainter than the point at which incompleteness sets in. Note that (1) the completeness 
curves for the two fields are very similar, and (2) the $\Delta K$ 
distributions of both fields are also similar, being skewed to 
positive values by only $\sim 0.1$ magnitudes. That the photometric properties 
of the two datasets are very similar, despite obvious differences in integrated 
surface brightness (e.g. Figure 1), indicates that the faint 
limit of the CFHTIR data is set by photon statistics, rather than crowding.}
\end{figure}

\begin{figure}
\figurenum{5}
\epsscale{0.95}
\plotone{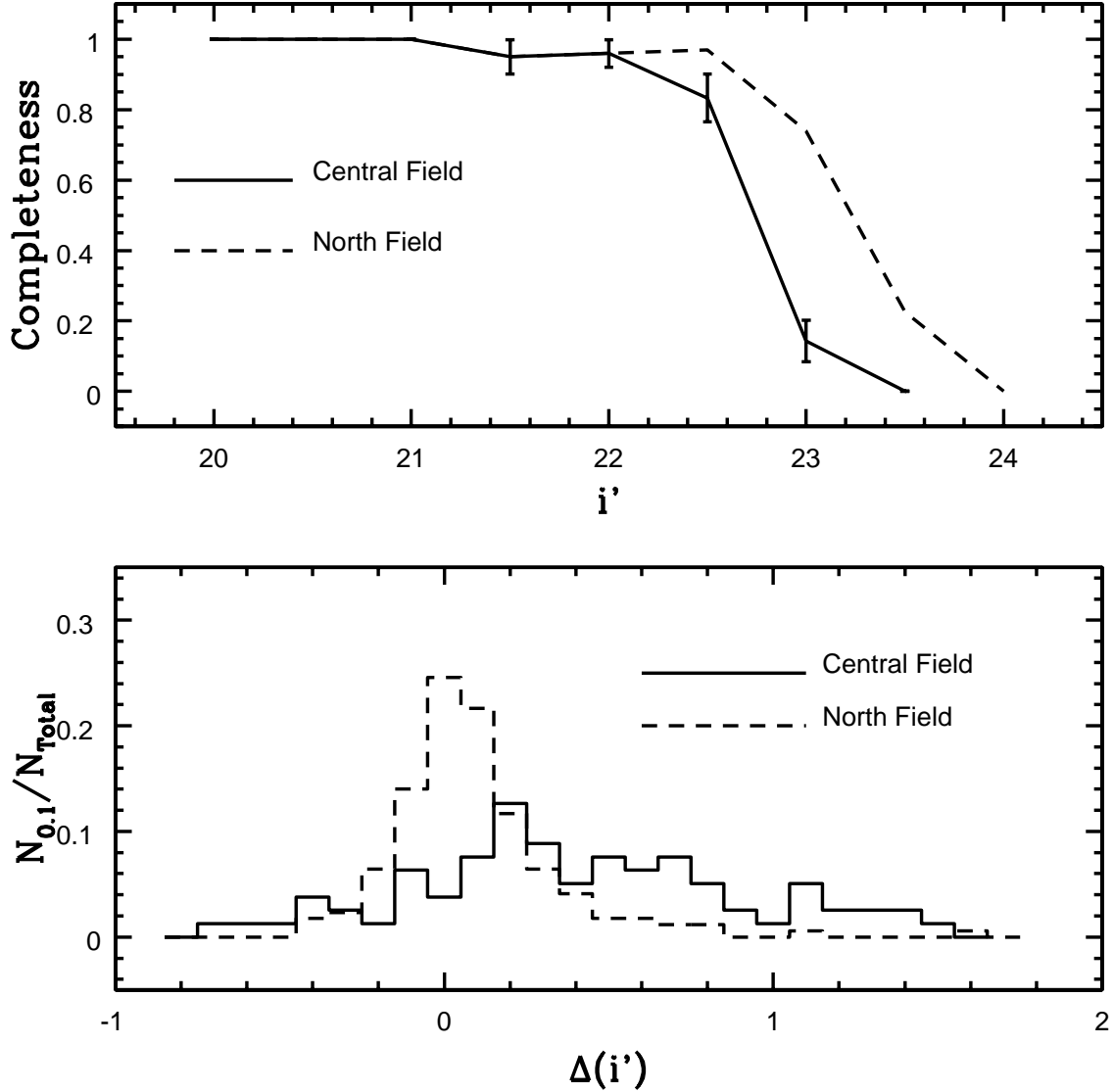}
\caption
{The results of artificial star experiments run on the regions of the MegaPrime data that 
coincide with the CFHTIR Center and North fields. The top panel shows the completeness 
fraction in $i'$, where completeness is the ratio of the number of artificial stars 
recovered to the number that were added. To be recovered, an artificial star had to be 
detected in both $r'$ and $i'$. Note that the completeness curve for the North field is 
0.5 magnitudes fainter than that of the Center field. 
The lower panel shows the difference between the actual and measured brightnesses 
of the artificial stars, $\Delta i'$, for $i' = 23$, which is 0.5 magnitudes fainter 
than the point at which incompleteness becomes significant in the central field. Note 
that the difference distribution for the Center field is very broad, and extends 
past $\Delta i' = 1$. Hence, crowding is a significant factor in setting the faint 
limit of the MegaPrime data near the center of NGC 247. 
For comparison, the $\Delta i'$ distribution of the 
North field is much narrower, indicating that blending is less of a problem in the 
lower surface brightness portions of the NGC 247 disk.}
\end{figure}

\begin{figure}
\figurenum{6}
\epsscale{1.0}
\plotone{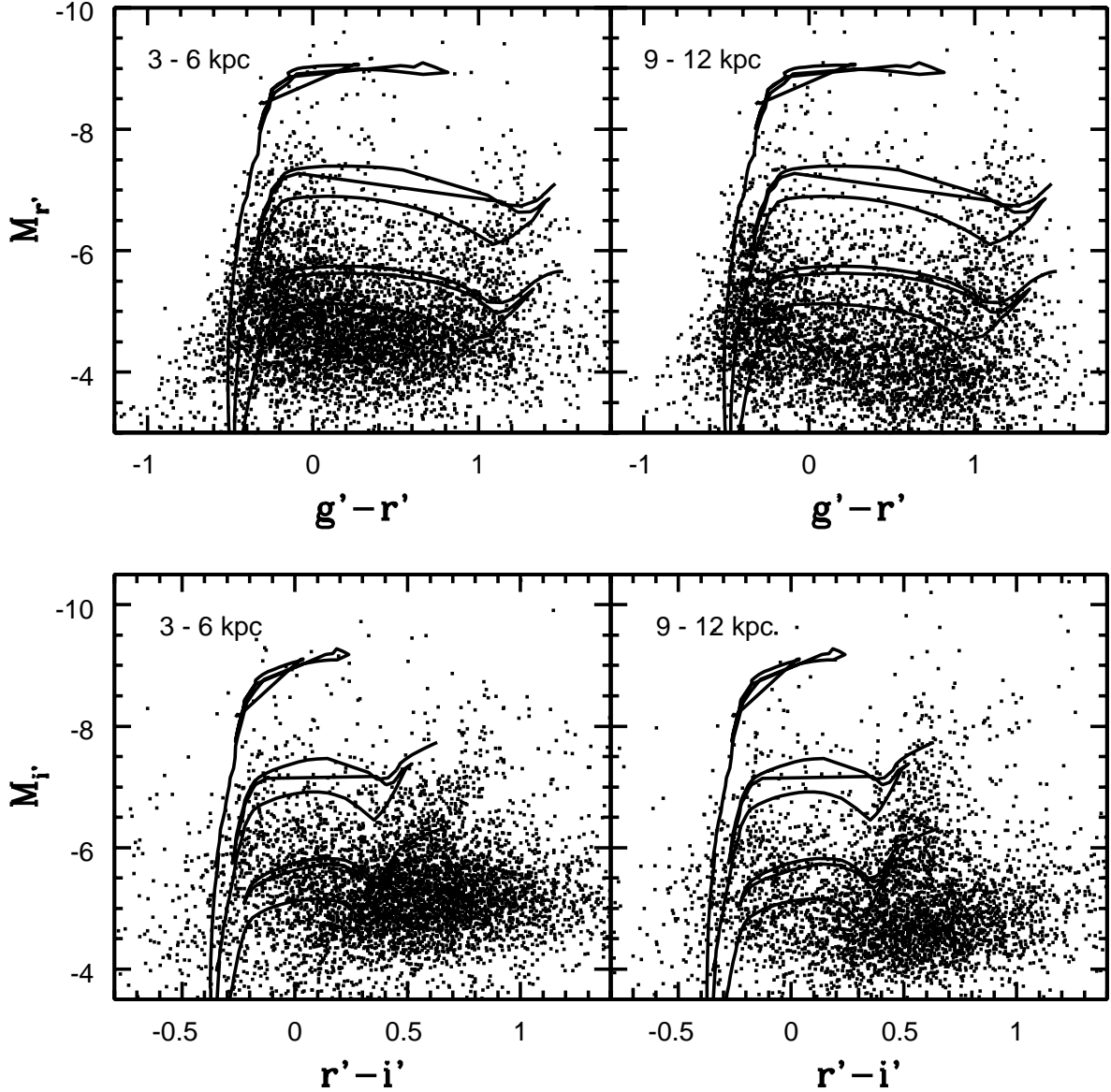}
\caption
{The $(M_{r'}, g'-r')$ and $(M_{i'}, r'-i')$ CMDs of sources in two radial intervals. 
The distances indicated in each panel are de-projected galactocentric radii. 
A distance modulus of 27.9 (\S 5), and a reddening A$_B = 0.46$, which 
accounts for internal and external sources of dust (Pierce \& Tully 1992), have 
been assumed. The isochrones are Z=0.008 sequences from Girardi et al. (2004)
with log(t$_{yr}$) = 6.8, 7.2, and 7.6.}
\end{figure}

\begin{figure}
\figurenum{7}
\epsscale{1.0}
\plotone{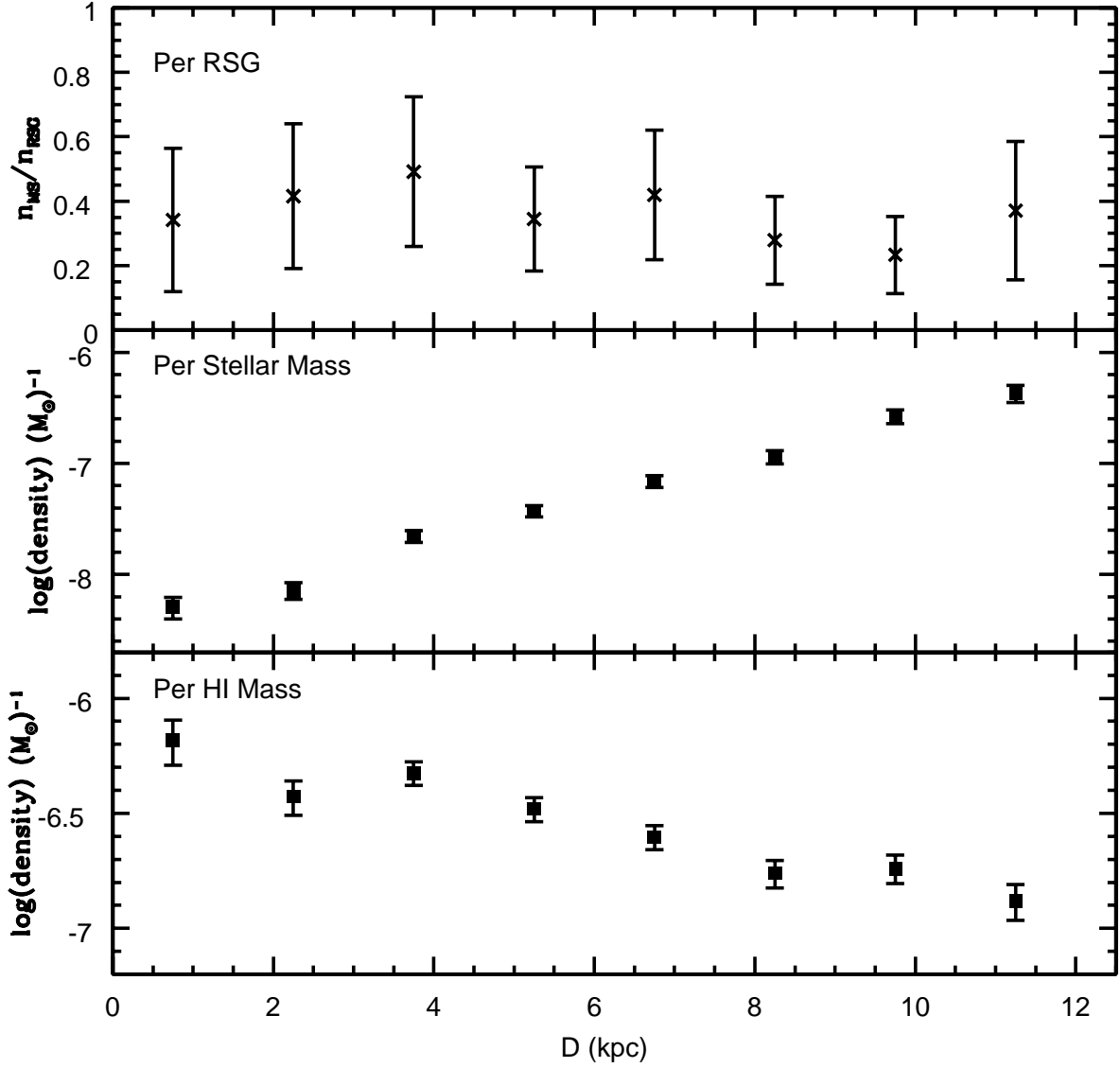}
\caption
{The number density of main sequence stars with M$_{i'}$ between 
--6.4 and --7.4, normalized to: (1) the number of RSGs with 
M$_{i'}$ in the same brightness interval (top panel), (2) integrated local stellar mass 
(middle panel), and (3) projected HI mass (bottom panel). The integrated local stellar 
mass in the middle panel was computed from the $K-$band surface 
brightness profile of NGC 247 in the 2MASS Extended Source Catalogue (Jarrett 
et al. 2000), assuming M/L$_K = 2$. The HI mass used to construct the lower 
panel was taken from the projected HI density profile shown in Figure 4 of 
Carignan \& Puche (1990), after being adjusted to a distance modulus of 27.9.} 
\end{figure}

\begin{figure}
\figurenum{8}
\epsscale{1.0}
\plotone{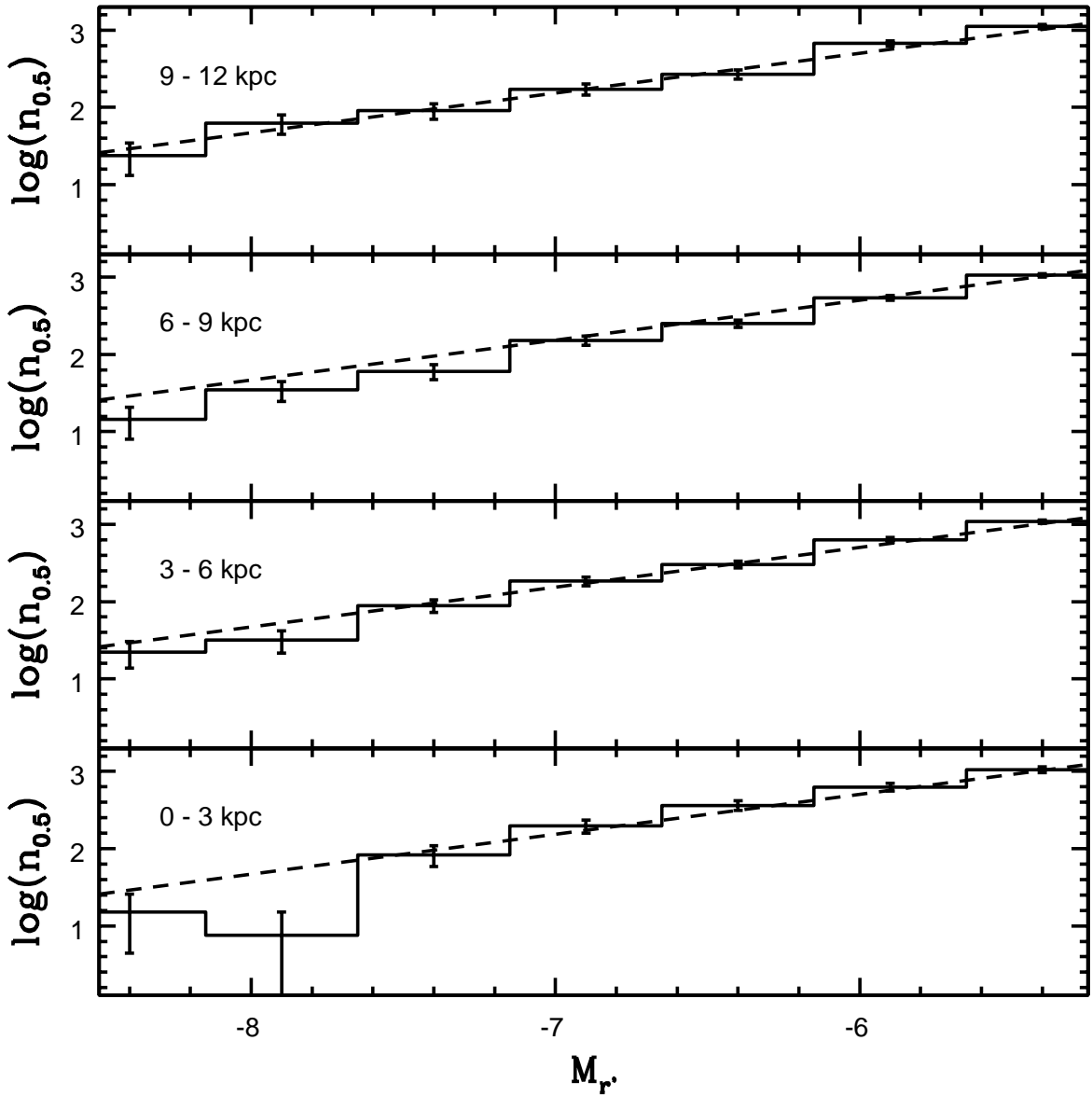}
\caption
{The M$_{r'}$ LFs of main sequence stars throughout the HI disk of NGC 247. The 
distances indicated in each panel refer to the de-projected radial intervals from which 
stars were extracted to construct each LF, while n$_{0.5}$ is the number of stars per 
0.5 magnitude interval in M$_{r'}$, scaled to match the total number in the NGC 247
disk with M$_{r'}$ between --6.1 and --4.6. The dashed line shows a least 
squares power-law that was fit to the LF of stars in all distance intervals with M$_{r'}$ 
between --7.1 and --4.6. A single power-law provides a reasonable match to 
the LFs in all distance intervals when M$_{r'} > -7$. However, there is an 
absence of stars with M$_{r'} < -7.5$ in the 0 -- 3 kpc annulus.}
\end{figure}

\begin{figure}
\figurenum{9}
\epsscale{1.0}
\plotone{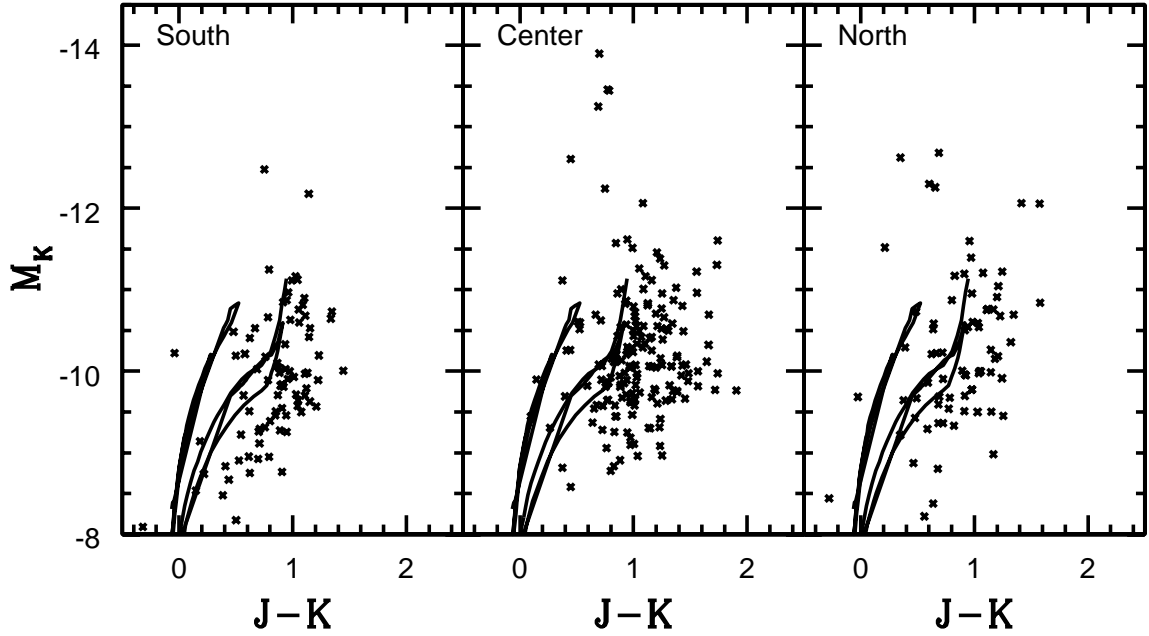}
\caption
{The $(M_K,J-K)$ CMDs of the three CFHTIR fields. A distance modulus of $\mu_0 = 
27.9$ has been adopted, based on the $i'$ brightness of the RGB-tip (\S 5). A combined 
internal and external reddening of A$_B = 0.46$ magnitudes (Pierce \& Tully 1992) 
has been assumed. The isochrones are Z=0.008 sequences from Girardi et al. (2002) 
with log(t$_{yr}$) = 6.8 and 7.0. Based on peak brightness, the models suggest 
that the brightest RSGs have log(t$_{yr}) \sim 7.0$, although the predicted RSG 
sequence with this age is $\sim 0.1$ magnitudes too blue.}
\end{figure}

\begin{figure}
\figurenum{10}
\epsscale{1.0}
\plotone{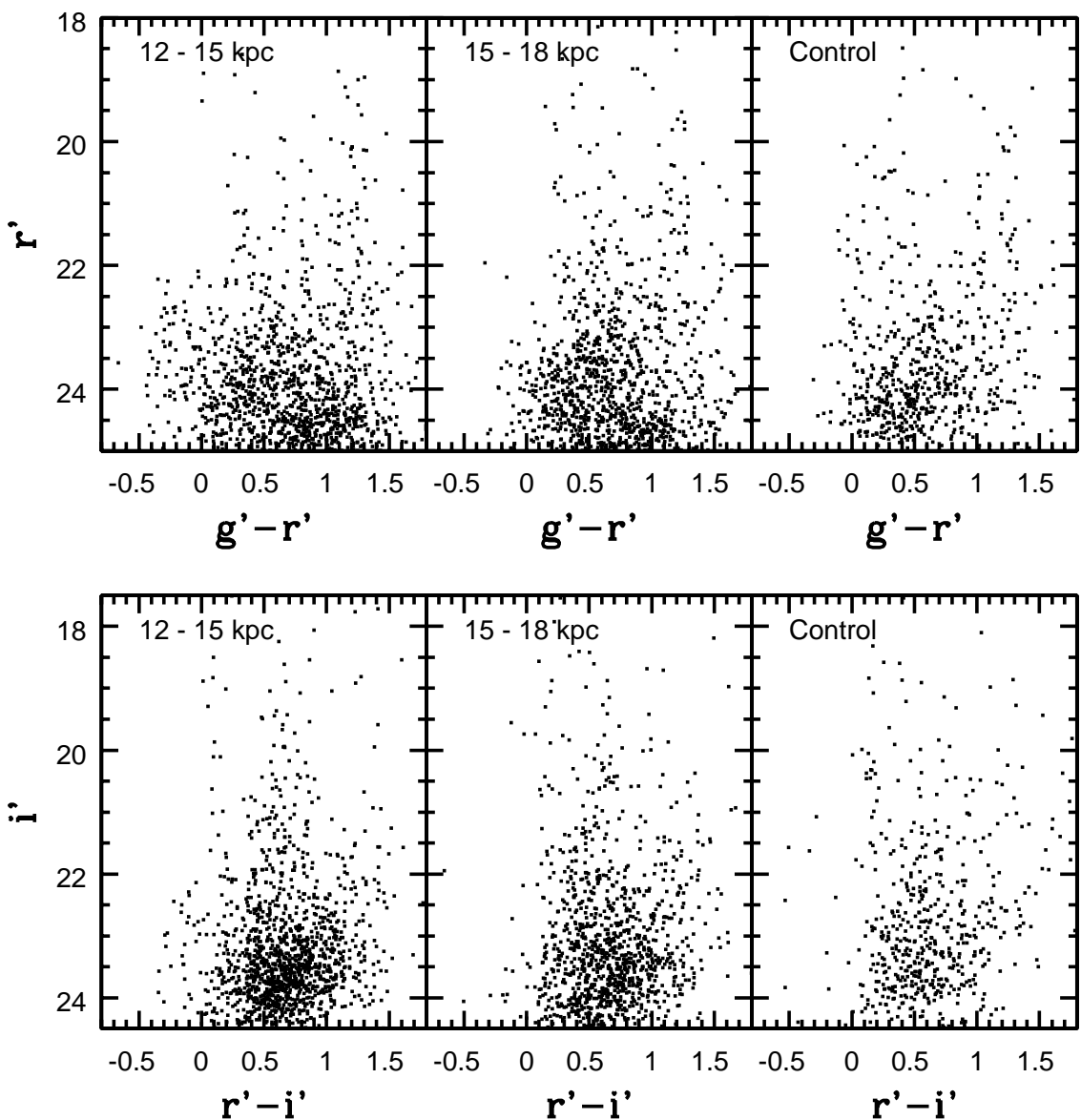}
\caption
{The $(r', g'-r')$ and $(i', r'-i')$ CMDs of stars in the 
outer disk of NGC 247. Also shown are the CMDs of a portion of the MegaPrime data that 
samples the foreground star and background galaxy populations, in a region having 
the same total area on the sky as the $15 - 18$ kpc annulus. Note that the main sequence 
and RSG plume are much more poorly defined than in Figure 2.}
\end{figure}

\begin{figure}
\figurenum{11}
\epsscale{1.0}
\plotone{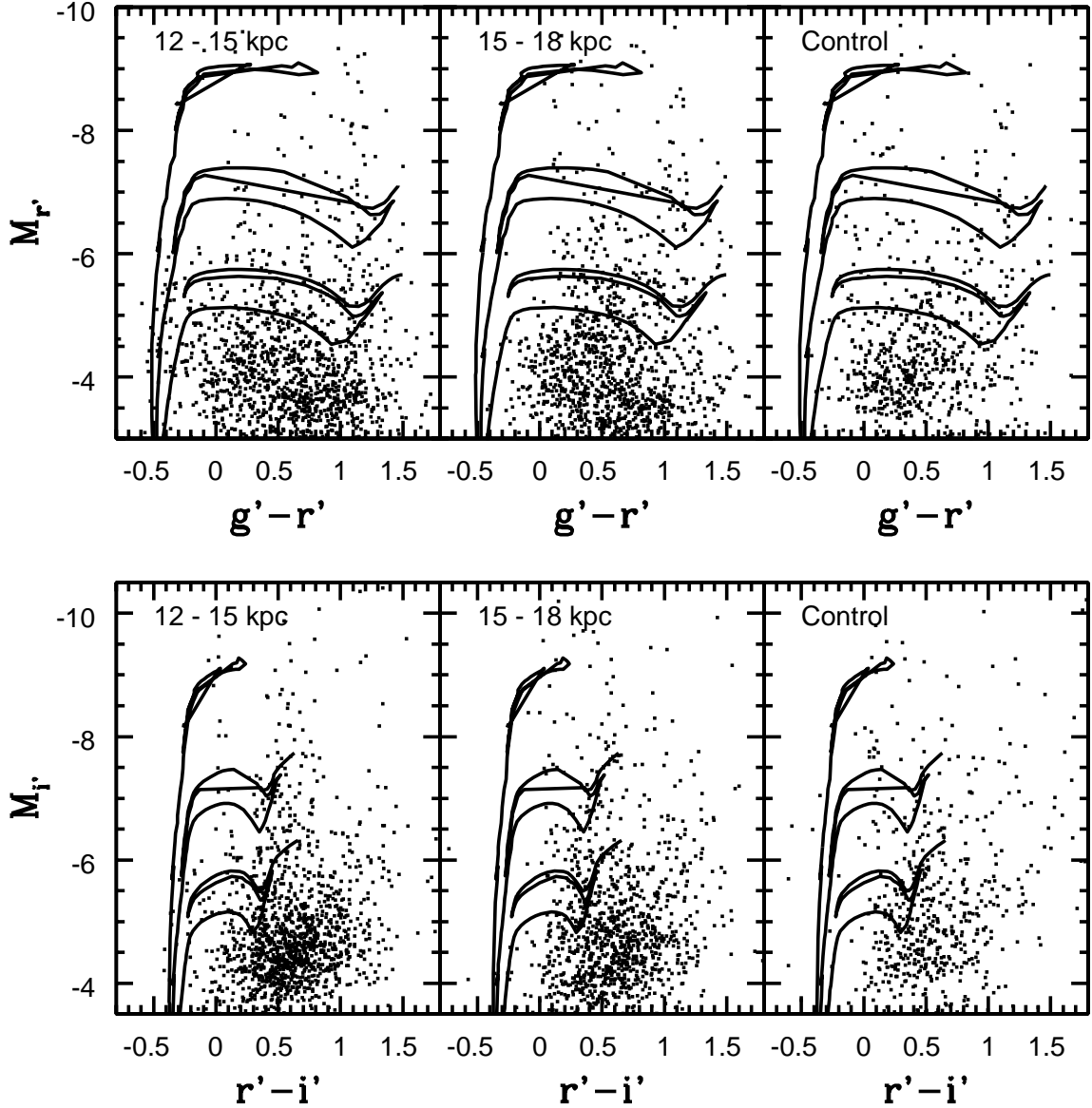}
\caption
{The $(M_{r'}, g'-r')$ and $(M_{i'}, r'-i')$ CMDs of stars in the outer 
disk of NGC 247. Also shown are CMDs of a portion of the MegaPrime data that samples the 
foreground star and background galaxy populations, extracted from a 
region having the same total area on the sky as the $15 - 18$ kpc annulus. 
A distance modulus of 27.9 and A$_B = 0.46$ magnitudes have been 
assumed. Isochrones from Girardi et al. (2004) with Z = 0.008 and log(t$_{yr}$) = 
6.8, 7.2, and 7.6 are also shown. Note that the main sequence turn-off in the 
$12 - 15$ kpc interval corresponds roughly to log(t$_{yr}$) = 7.2, while in the $15 - 18$ 
kpc interval the brightest main sequence stars have log(t$_{yr}) > 7.6$.}
\end{figure}

\begin{figure}
\figurenum{12}
\epsscale{1.0}
\plotone{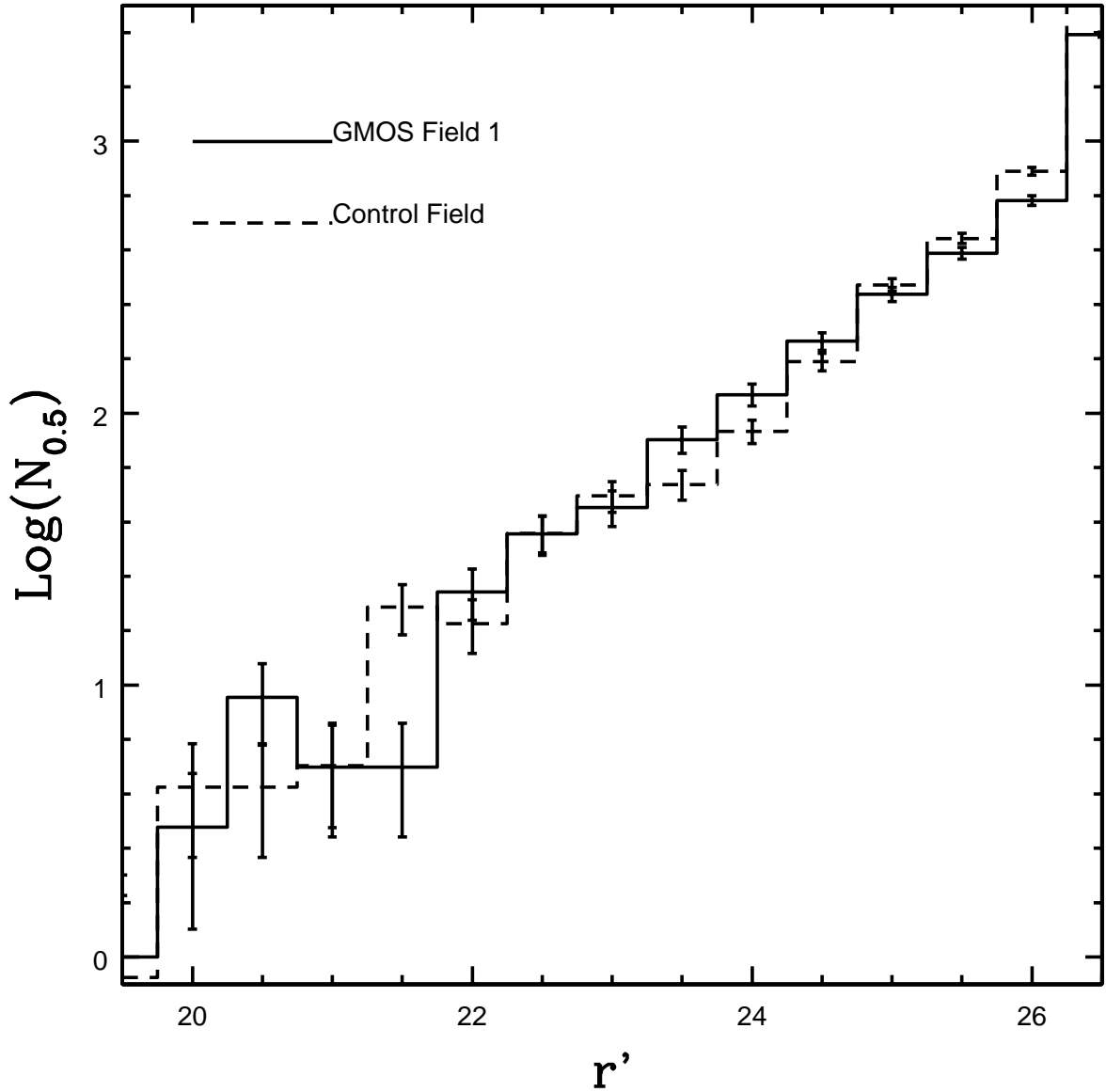}
\caption
{The $r'$ LFs of sources in GMOS Field 1 and the control field. N$_{0.5}$ 
is the number of stars per 0.5 magnitude interval in GMOS Field 1. The error 
bars show the uncertainties due to counting statistics. The 
good agreement between the two LFs indicates that the source counts in Field 1 
are dominated by foreground stars and background galaxies.}
\end{figure}

\begin{figure}
\figurenum{13}
\epsscale{1.0}
\plotone{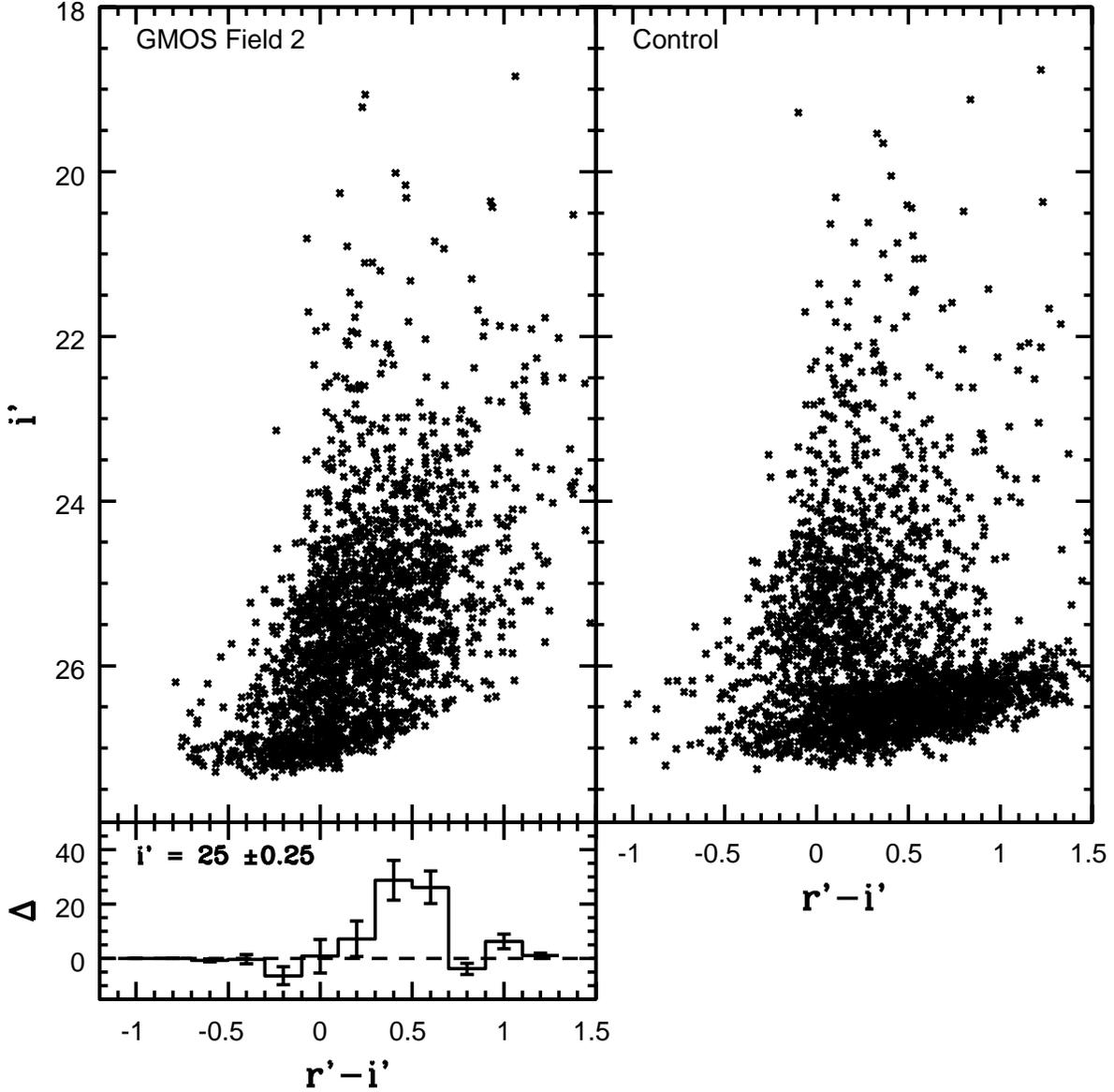}
\caption
{The $(i', r'-i')$ CMDs of GMOS Field 2 and the control field. Note the 
excess population of objects with $r'-i' > 0.1$ and $i' > 24$ in GMOS 
Field 2 when compared with the control field; 
there are 571 stars in GMOS Field 2 with $i'$ between 24 and 25.5 and 
$r'-i'$ between 0 and 1, while there are 363 objects in the corresponding portion of 
the control field CMD. This excess population in GMOS Field 2 is made up of RGB stars in
NGC 247, and the difference between the number of objects with $i'$ between 24.75 and 
25.25 in GMOS Field 2 and the control field is shown per 0.2 magnitude $r'-i'$ interval in 
the lower left hand panel. Note that a significant excess number of objects 
are seen in GMOS Field 2 when $r'-i'$ is between 0.3 and 0.7.} 
\end{figure}

\begin{figure}
\figurenum{14}
\epsscale{1.0}
\plotone{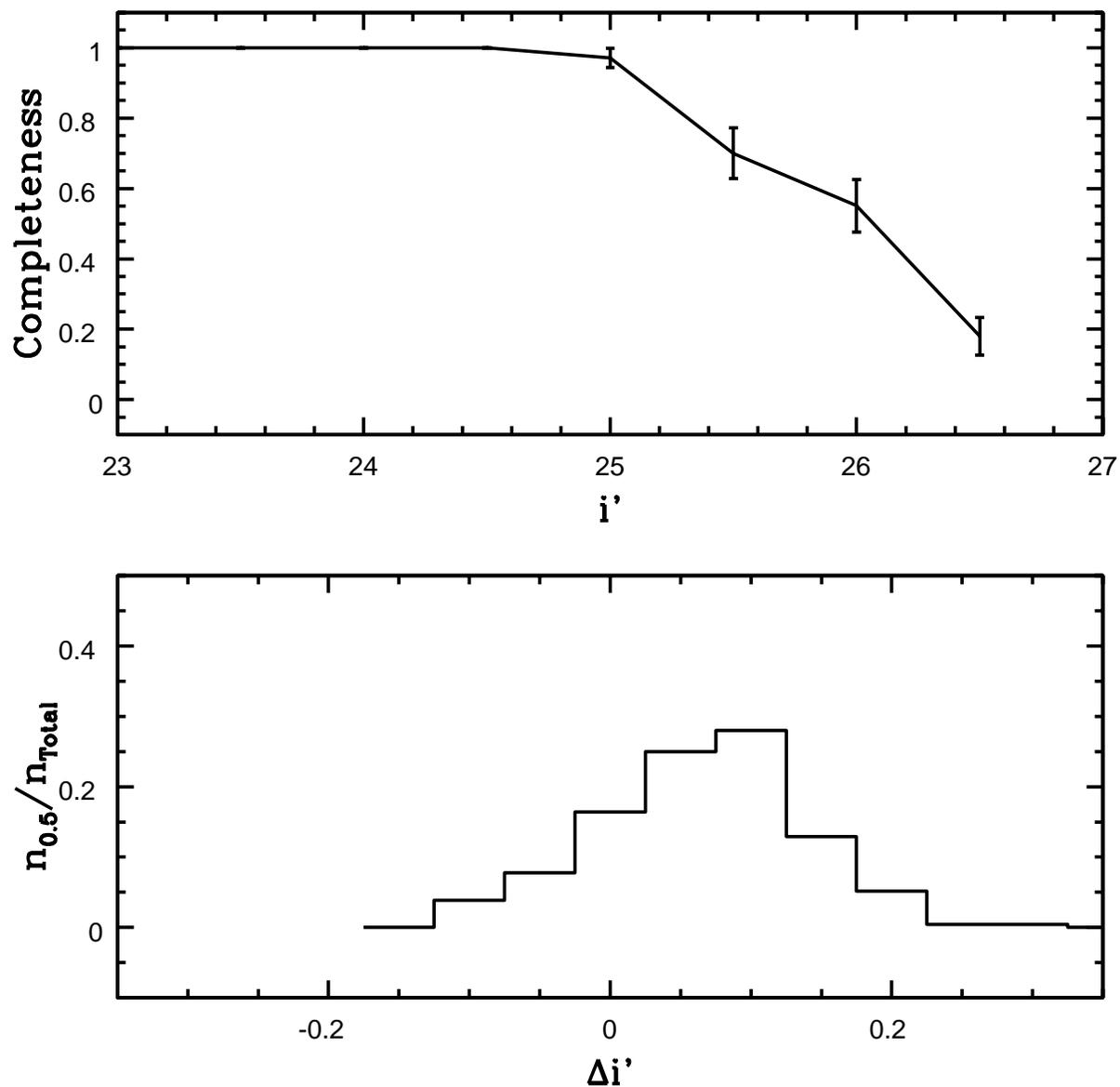}
\caption
{The results of artificial star experiments run on the GMOS Field 2 
observations. The top panel shows the completeness fraction in $i'$, where completeness 
is the number of artificial stars detected in both $r'$ and $i'$ divided by 
the number that were added. The lower panel shows the difference between the actual and 
measured brightnesses of artificial stars with $i' = 25$, which is the magnitude 
at which incompleteness sets in. The distribution has a standard deviation of 
$\pm 0.1$ mag. The absence of recovered stars with magnitudes that 
are vastly brighter than their actual values 
suggests that blending is not an issue with these data.}
\end{figure}

\begin{figure}
\figurenum{15}
\epsscale{1.0}
\plotone{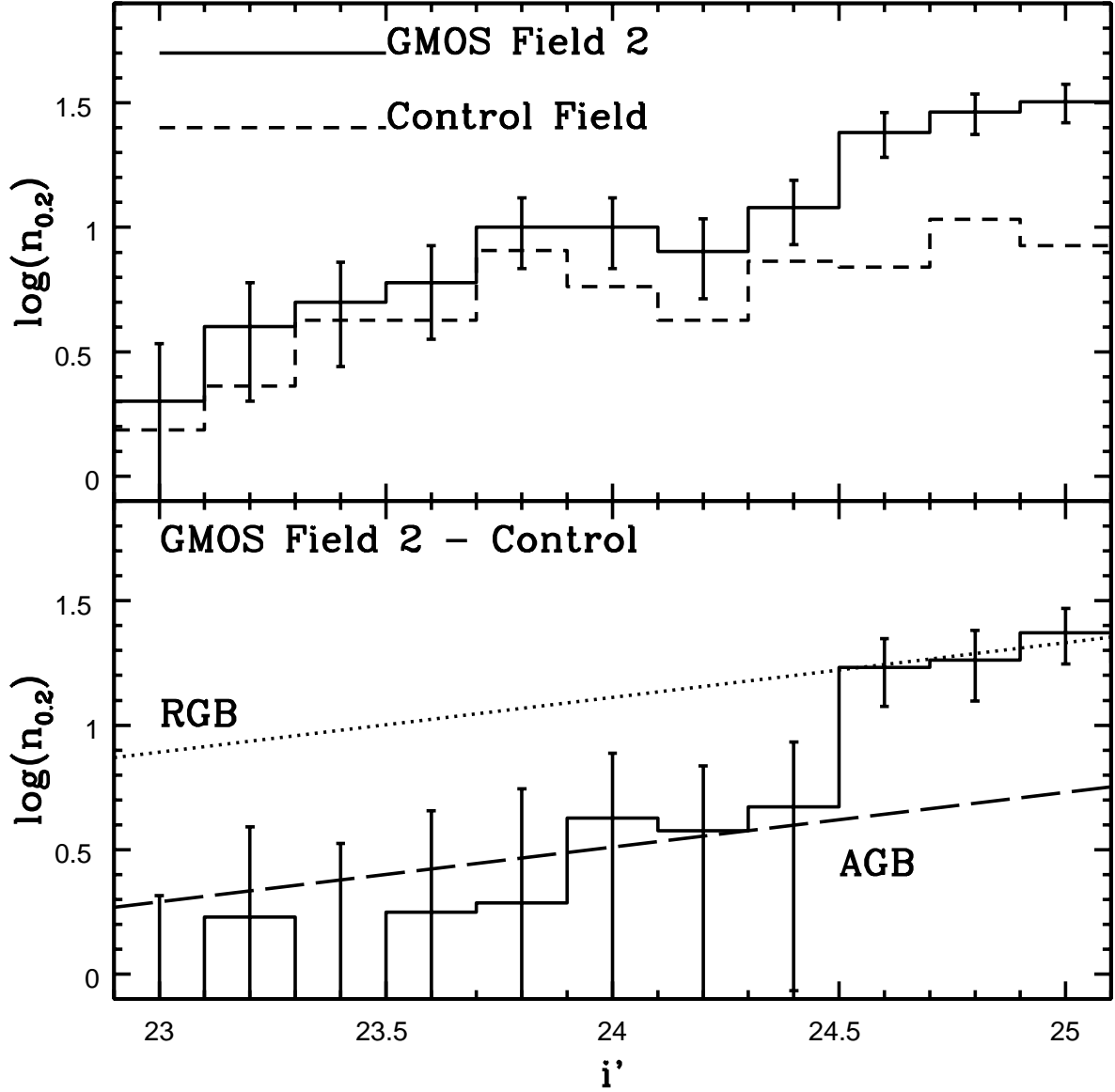}
\caption
{The $i'$ LFs of the half of GMOS Field 2 that is closest to NGC 247 and the control 
field are compared in the upper panel. The control field LF has been scaled to account for 
the smaller size of the GMOS Field 2 region, and n$_{0.2}$ is the number of sources per 
0.2 magnitude $i'$ interval in GMOS Field 2 with $r'-i'$ between 0.3 and 0.9. The 
difference between the LFs is shown in the lower panel, and a discontinuity can be seen 
near $i' = 24.5$. The dotted line in the lower panel shows a least squares 
fit of an $x = 0.22$ power-law to the entries with $i' > 24.5$, while the 
dashed line shows this same relation, but shifted down assuming that the ratio of 
RGB to AGB stars is $5:1$. Note that the expected AGB relation 
roughly passes through the entries in the differenced LF at the bright end. Based on the 
comparisons in this figure, it is concluded that the RGB-tip in NGC 247 
occurs at $i' = 24.5 \pm 0.1$.}
\end{figure}

\begin{figure}
\figurenum{16}
\epsscale{1.0}
\plotone{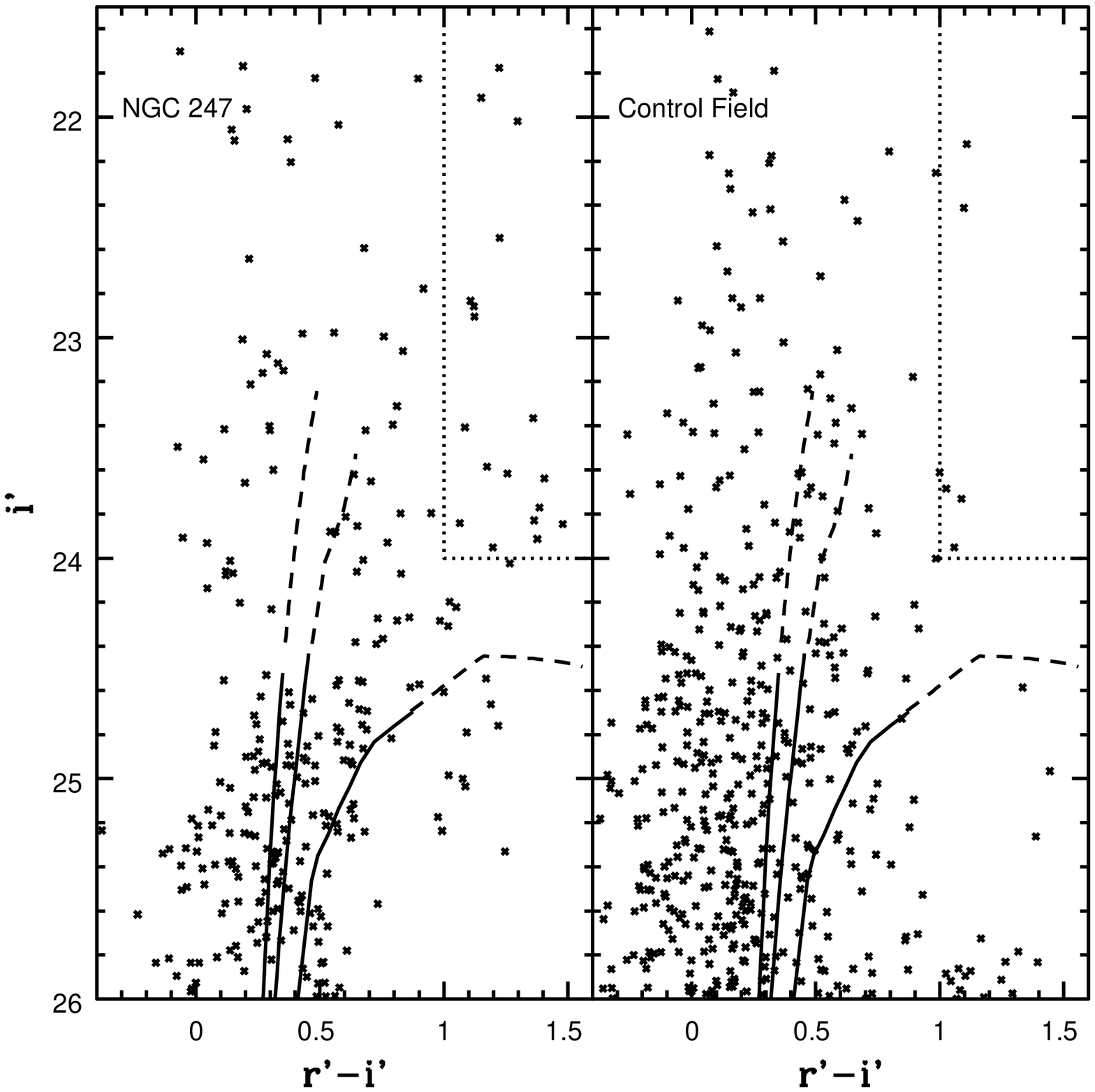}
\caption
{The left hand panel shows the $(i', r'-i')$ CMD of the half of GMOS Field 2 that is 
closest to NGC 247, after sources that have been paired with objects in the 
control field using the procedure described in the text have been deleted. 
The right hand panel shows the $(i', r'-i')$ CMD of 
a portion of the control field that has the same area as one half of GMOS Field 2. The 
solid lines are Z=0.0001, 0.001, and 0.004 RGB sequences with log(t$_{yr}$) = 10.0 from 
Girardi et al. (2005), assuming a distance modulus of 27.9. The dashed lines show the 
upper AGB sequences from these models. The dotted lines mark the area with $r'-i' < 
1$ and $i' < 24$, which is where AGB stars with log(t$_{yr}) < 10$ will occur. Note that 
the number of objects in this region of the GMOS Field 2 CMD far exceeds those 
in the control field, suggesting that they are likely not 
a fluke population of background sources.}
\end{figure}


\clearpage

\begin{references}

\reference{} Barnes, J., \& Efstathiou, G. 1987, ApJ, 319, 575

\reference{} Battistini, P., B\`{o}noli, F., Federici, L., Fusi Pecci, F., \& Kron, R. G. 1984, A\&A, 130, 162

\reference{} Bekki, K., \& Chiba, M. 2001, ApJ, 558, 666

\reference{} Bekki, K., \& Freeman, K. C. 2003, MNRAS, 346, L11

\reference{} Bell, E. F., \& de Jong, R. S. 2000, MNRAS, 312, 497

\reference{} Bellazzini, M., Ferraro, F. R., \& Ibata, R. 2003, AJ, 125, 188

\reference{} Beasley, M. A., Brodie, J. P., Strader, J., Forbes, D. A., Proctor, R. N., Barmby, P., \& Huchra, J. P. 2004, AJ, 128, 1623

\reference{} Bensby, T., Feltzing, S., \& Lundstrom, I. 2003, A\&A, 410, 527

\reference{} Bensby, T., Feltzing, S., \& Lundstrom, I. 2004, A\&A, 421, 969

\reference{} Binney, J., Dehnen, W., \& Bertelli, G. 2000, MNRAS, 318, 658

\reference{} Bland-Hawthorn, J., Vlajic, M., Freeman, K. C., \& Draine, B. T. 2005, ApJ, 629, 239

\reference{} Boulade, O. et al. 2003, Proc. SPIE, 4841, 72

\reference{} Brooks, R. S., Wilson, C. D., \& Harris, W. E. 2004, AJ, 128, 237

\reference{} Brown, T. M., et al. 2003, ApJ, 592, L17

\reference{} Butler, D. J., Martinez-Delgado, D., \& Brandner, W. 2004, AJ, 127, 1472

\reference{} Cardelli, J. A., Clayton, G. C., \& Mathis, J. S. 1989, ApJ, 345, 245

\reference{} Carignan, C. 1985, ApJS, 58, 107

\reference{} Carignan, C., \& Puche, D. 1990, AJ, 100, 641

\reference{} Condon, J. J. 1987, ApJS, 65, 485

\reference{} Condon, J. J. 1992, ARA\&A, 30, 575

\reference{} Crampton, D., et al. 2000, Proc. SPIE, 4008, 114

\reference{} Cutri, R. M. et al. 2003, 2MASS All-Sky Catalogue of Point Sources (Amherst: Univ. Massachusetts Press; Pasadena: IPAC)

\reference{} Dalcanton, J. J., \& Bernstein, R. A. 2002, AJ, 124, 1328

\reference{} Davidge, T. J. 1998, ApJ, 497, 650

\reference{} Davidge, T. J. 2000, AJ, 119, 748

\reference{} Davidge, T. J. 2003, AJ, 125, 3046

\reference{} Davidge, T. J. 2005, ApJ, 622, 279

\reference{} Davidge, T. J., \& Courteau, S. 2002, AJ, 123, 1438

\reference{} de Grijs, R., Kregel, M., \& Wesson, K. H. 2001, MNRAS, 324, 1074

\reference{} de Vaucouleurs, G. 1978, ApJ, 224, 710

\reference{} de Vaucouleurs, G. 1979, ApJ, 227, 729

\reference{} de Vaucouleurs, G., de Vaucouleurs, A., Corwin, H. G., Buta, R., Paturel, G., \& Forque, P. 1991, Third Reference Catalogue of Bright Galaxies (Austin: University of Texas Press).

\reference{} Eggenberger, P., Meynet, G., \& Maeder, A. 2002, A\&A, 386, 576

\reference{} Erwin, P., Beckman, J. E., \& Pohlen, M. 2005, ApJ, 626, L81

\reference{} Ferguson, A. M. N., \& Johnson, R. A. 2001, ApJ, 559, L13

\reference{} Ferguson, A. M. N., Wyse, R. F. G., Gallagher, J. S. III, \& Hunter, D. A. 1996, AJ, 111, 2265

\reference{} Freedman, W. L. 1985, ApJ, 299, 74

\reference{} Gallart, C., Stetson, P. B., Hardy, E., Pont, F., \& Zinn, R. 2004, ApJ, 614, L109

\reference{} Galleti, S., Bellazzini, M., \& Ferraro, F. R. 2004, A\&A, 423, 925

\reference{} Gilmore, G., Wyse, R. F. G., \& Norris, J. E. 2002, ApJ, 574, L39

\reference{} Girardi, L., Bertelli, G., Bressan, A., Chiosi, C., Groenewegen, M. A. T., Marigo, P., Salasnich, B., \& Weiss, A. 2002, A\&A, 391, 195

\reference{} Girardi, L., Grebel, E. K., Odenkirchen, M., \& Chiosi, C. 2004, A\&A, 422, 205

\reference{} Gordon, K. D., Hanson, M. M., Clayton, G. C., Rieke, G. H., \& Misselt, K. A. 1999, ApJ, 519, 165

\reference{} Governato et al. 2004, ApJ, 607, 688

\reference{} Hawarden, T. G., Leggett, S. K., Letawsky, M. B., Ballantyne, D. R., \& Casali, M. M. 2001, MNRAS, 325, 563

\reference{} Haynes, M. P., \& Roberts, M. S. 1979, ApJ, 227, 767

\reference{} Humphreys, R. M., \& McElroy, D. B. 1984, ApJ, 284, 565

\reference{} Ibata, R., Irwin, M., Lewis, G. F., \& Stolte, A. 2001, ApJ, 547, L133

\reference{} Ibata, R. A., Irwin, M. J., Lewis, G. F., Ferguson, A. M. N., \& Tanvir, N. 2003, MNRAS, 340, L21

\reference{} Ibata, R., Chapman, S., Ferguson, A. M. N., Irwin, M., Lewis, G., \& McConnachie, A. 2004, MNRAS, 351, 117

\reference{} Jarrett, T. H., Chester, T., Cutri, R., Schneider, S., Skrutskie, M., \& Huchra, J. P. 2000, AJ, 119, 2498

\reference{} Karachentsev, I. D. et al. 2003, A\&A, 404, 93

\reference{} Kennicutt, R. C. 1989, ApJ, 344, 685

\reference{} Labbe, I., et al. 2003, ApJ, 591, L95

\reference{} Landolt, A. U. 1992, AJ, 104, 340

\reference{} Langer, N., \& Maeder, A. 1995, A\&A, 295, 685

\reference{} Liu, W. M., \& Chaboyer, B. 2000, ApJ, 544, 818

\reference{} Marleau, F. R., \& Simard, L. 1998, ApJ, 507, 585

\reference{} Martin, C. L., \& Kennicutt Jr., R. C. 2001, ApJ, 555, 301

\reference{} Martinez-Delgado, D., Aparicio, A., Gomez-Flechoso, M. A., \& Carrera, R. 2001, ApJ, 549, L199

\reference{} Mayer, L., et al. 2001, ApJ, 559, 754

\reference{} McLean, I. S., \& Liu, T. 1996, ApJ, 456, 499

\reference{} Miller, G. E., \& Scalo, J. M. 1979, ApJS, 41, 513

\reference{} Mizutani, A., Chiba, M., \& Sakamoto, T. 2003, ApJ, 589, L89

\reference{} Moore, B., Lake, G., \& Katz, N. 1998, ApJ, 495, 139

\reference{} Morrison, H. L., Harding, P., Perrett, K., \& Hurley-Keller, D. 2004, ApJ, 603, 87

\reference{} Mouhcine, M., \& Lancon, A. 2003, A\&A, 402, 425

\reference{} Narayan, C. A., \& Jog, C. J. 2003, A\&A, 407, L59

\reference{} Navarro, J. F., \& Steinmetz, M. 1997, ApJ, 478, 13

\reference{} Nishiyama, K., Nakai, N., \& Kuno, N. 2001, PASJ, 53, 757

\reference{} O'Donnell, J. E. 1994, ApJ, 422, 1580

\reference{} Olsen, K. A. G., Miller, B. W., Suntzeff, N. B., Schommer, R. A., \& Bright, J. 2004, AJ, 127, 2674

\reference{} Pasetto, S., Chiosi, C., \& Carraro, G. 2003, A\&A, 405, 931

\reference{} Pierce, M. J., \& Tully, R. B. 1992, ApJ, 387, 47

\reference{} Putman, M. E., Staveley-Smith, L., Freeman, K. C., Gibson, B. K., \& Barnes, D. G. 2003, ApJ, 586, 170

\reference{} Quinn, P. J., \& Goodman, J. 1986, ApJ, 309, 472

\reference{} Read, A. M., \& Ponman, T. J. 2001, MNRAS, 328, 127

\reference{} Rice, W. et al. 1988, ApJS, 68, 91

\reference{} Robin, A. C., Reyle, C., Derriere, S., \& Picaud, S. 2003, A\&A, 409, 523

\reference{} Rossa, J., Dettmar, R-J, Walterbos, R. A. M., \& Norman, C. A. 2004, AJ, 128, 674

\reference{} Rozanski, R., \& Rowan-Robinson, M. 1994, MNRAS, 271, 530

\reference{} Salaris, M., Weiss, A., \& Percival, S. M. 2004, A\&A, 414, 163

\reference{} Sandage, A., \& Tammann, G. A. 1987, A Revised Shapley-Ames Catalog of Bright Galaxies, (Washington: Carnegie Institute of Washington).

\reference{} Sarajedini, A., Geisler, D., Harding, P., \& Schommer, R. 1998, 508, L37

\reference{} Schommer, R. A., Suntzeff, N. B., Olszewski, E. W., \& Harris, H. C. 1992, AJ, 103, 447

\reference{} Schlegel, D. J., Finkbeiner, D. P., \& Davis, M. 1998, ApJ, 500, 525

\reference{} Searle, L., \& Zinn, R. 1978, ApJ, 225, 357

\reference{} Sellwood, J. A., \& Binney, J. J. 2003, MNRAS, 336, 785

\reference{} Stephens, A. W., \& Frogel, J. A. 2002, AJ, 124, 2023

\reference{} Stetson, P. B. 1987, PASP, 99, 191

\reference{} Stetson, P. B., \& Harris, W. E. 1988, AJ, 96, 909

\reference{} Strassle, M., Huser, M., Jetzer, P., \& De Paolis, F. 1999, A\&A, 349, 1

\reference{} Thilker, D. A., et al. 2004, ApJ, 601, L39

\reference{} Tiede, G. P., Sarajedini, A., \& Barker, M. K. 2004, AJ, 128, 224

\reference{} Tsuchiya, T., Dinescu, D. I., \& Korchagin, V. I. 2003, ApJ, 589, L29

\reference{} Tullmann, R., Rosa, M. R., Elwert, T., Bomans, D. J., Ferguson, A. M. N., \& Dettmar, R.-J. 2003, A\&A, 412, 69

\reference{} Tully, R. B., \& Fouque, P. 1985, ApJS, 58, 67

\reference{} van der Kruit, P. C., \& Searle, L. 1981, A\&A, 95, 105

\reference{} Warmels, R. H. 1988, A\&AS, 72, 19

\reference{} Weil, M. L., Eke, V. R., \& Efstathiou, G. 1998, MNRAS, 300, 773

\reference{} White, S. D. M. 1984, ApJ, 286, 38

\reference{} Whiting, A. B. 1999, AJ, 117, 202

\reference{} Zaritsky, D., Elston, R., \& Hill, J. M. 1989, AJ, 97, 97

\reference{} Zibetti, S., \& Ferguson, A. M. N. 2004, MNRAS, 352, L6

\end{references}
\end{document}